\documentclass[pre,aps,reprint,longbibliography]{revtex4-1}

\usepackage{amsmath}
\usepackage{amssymb}
\usepackage{graphicx}
\usepackage{subfigure}
\usepackage{bm}
\usepackage{color}
\usepackage{commath}
\usepackage{dcolumn}
\usepackage{multirow}

\usepackage{hyperref} 

\begin{document}

\title{Circumventing spin glass traps by microcanonical spontaneous symmetry breaking}

\author{Hai-Jun Zhou$^{1,2,3,4}$}
\thanks{Email: zhouhj@itp.ac.cn}
\author{Qinyi Liao$^{1}$}
\thanks{Email: qinyi.liao.phy@gmail.com}

\affiliation{
  $^1$CAS Key Laboratory for Theoretical Physics, Institute of Theoretical Physics, Chinese Academy of Sciences, Beijing 100190, China}

\affiliation{
  $^2$School of Physical Sciences, University of Chinese Academy of Sciences, Beijing 100049, China}

\affiliation{
  $^3$MinJiang Collaborative Center for Theoretical Physics, MinJiang University, Fuzhou 350108, China}

\affiliation{
  $^4$Synergetic Innovation Center for Quantum Effects and Applications, Hunan Normal University, Changsha 410081, China}

\date{\today}

\begin{abstract}
  The planted $p$-spin interaction model is a paradigm of random-graph systems possessing both a ferromagnetic phase and a disordered phase with the latter splitting into many spin glass states at low temperatures. Conventional simulated annealing dynamics is easily blocked by these low-energy spin glass states. Here we demonstrate that, actually this planted system is exponentially dominated by a microcanonical polarized phase at intermediate energy densities. There is a discontinuous microcanonical spontaneous symmetry breaking transition from the paramagnetic phase to the microcanonical polarized phase. This transition can serve as a mechanism to avoid all the spin glass traps, and it is accelerated by the restart strategy of microcanonical random walk. We also propose an unsupervised learning problem on microcanonically sampled configurations for inferring the planted ground state.
\end{abstract}

\maketitle

\section{Introduction}

The planted $p$-spin interaction model on a finite-connectivity random graph is a representative ferromagnetic system with many coexisting spin glass states. This model system has played an important role in understanding the physics of structural glasses~\cite{Franz-etal-2001,Montanari-RicciTersenghi-2004}. This system is also quite relevant and challenging to the field of statistical inference~\cite{Zdeborova-Krzakala-2016,Mezard-Montanari-2009}. It is equivalent to the generalized Sourlas code problem in information science~\cite{Sourlas-1989,Huang-Zhou-2009} and the planted maximum XORSAT problem in computer science~\cite{Watanabe-2013}. The large number of low-energy disordered spin glass configurations pose a huge computational challenge to retrieve the planted ground state.

When the environmental temperature slowly decreases, as occurs in a simulated annealing (SA) dynamics~\cite{Kirkpatrick-etal-1983}, the system is predicted to experience an equilibrium phase transition from the disordered paramagnetic phase to the ordered ferromagnetic phase. But this equilibrium transition is extremely difficult to occur in practice when the interactions are many-body in nature ($p\! \geq \! 3$), as the free-energy barrier between these two phases is proportional to the system size $N$ and the crystal-nucleation mechanism then fails. The system instead remains in the paramagnetic phase as temperature further decreases~\cite{Franz-etal-2001,Montanari-RicciTersenghi-2004}.

\begin{figure}
  \centering
  \includegraphics[angle=270, width=0.4\textwidth]{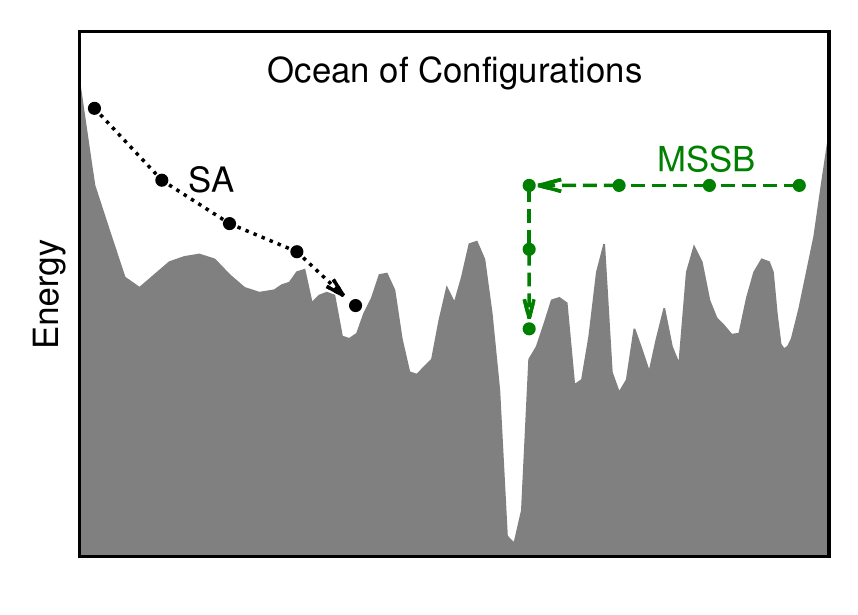}
  \caption{
    \label{fig:Landscape}
    Schematic energy landscape of the planted $p$-spin interaction system of size $N$ ($p \geq 3$). The white region contains all the $2^N$ microscopic configurations, the ``ocean of configurations''. An exponential number of deep local minima (the spin glass states) exist at the bottom of the energy landscape, which may trap the simulated annealing dynamics. We propose to take advantage of the discontinuous microscopic spontaneous symmetry breaking that occurs at high energies to directly jump to the basin of attraction of the ground state without encountering spin glass configurations.
  }
\end{figure}

In typical SA simulations with a small but finite rate of temperature decreasing, the system will decrease its energy with an exceedingly smaller rate, and finally it will reach one of the many spin glass states at the bottom of the energy landscape (Fig.~\ref{fig:Landscape}) and be trapped there for an exponentially long time (in $N$).  Besides, we have checked that message-passing algorithms based on belief propagation also fail to retrieve the ferromagnetic ground state. Following the proposal of Ref.~\cite{Xu-etal-2018}, we clamped the objective energy to a value much below those of the spin glass phase to search for ferromagnetic states, but we found that the message-passing iteration process does not converge to the ferromagnetic fixed point. This is because the high gauge symmetry of the planted system makes it statistically indistinguishable from the unplanted purely disordered system~\cite{Matsuda-etal-2011,Bandeira-etal-2018}.

This tremendous computational difficulty is not specific only to the planted $p$-spin system but is a common property of many large inference problems such as the planted satisfiability and coloring problems~\cite{Li-Ma-Zhou-2009,Krzakala-Mezard-Zdeborova-2014,Krzakala-Zdeborova-2009}. When the ground state is completely masked by an exponential number of disordered spin glass configurations (for example through quiet planting as in the $p$-spin system and the coloring problem~\cite{Krzakala-Zdeborova-2009}), it becomes evident that searching by SA and other dynamical processes at the bottom of the energy landscape and at low temperatures is not a promising strategy.

In this work, we demonstrate the benefit of working at sufficiently high energy values. Especially we point out the existence of an alternative route to the planted ground state, the route of microcanonical spontaneous symmetry breaking (MSSB). This is a discontinuous transition from the paramagnetic (or disordered symmetric, DS) phase to the microcanonical polarized (MP) phase. Because the energy density of the system along the whole DS-to-MP transition trajectory is fixed at a high value much above those of the spin glass states, the difficulty of climbing huge energy barriers between the different spin glass states is completely avoided (Fig.~\ref{fig:Landscape}). The microcanonically stable MP phase has so far been largely overlooked in the literature (except for a recent detailed analysis concerning the Potts model~\cite{Zhou-2019}), but we find that it actually is exponentially more dominant than the paramagnetic phase at intermediate energy density values.

The DS and MP phases are connected by many transition trajectories at energy densities $u \! \in \! (u_{\rm d}, u_{\rm mic})$, where $u_{\rm d}$ is the spin glass ergodicity breaking point and $u_{\rm mic}$ is the critical energy density of the equilibrium MSSB phase transition. As a main result of this work,  we find that $u_{\rm mic} \! > \! u_{\rm d}$.  Since ergodicity is preserved for $u \! > \! u_{\rm d}$ and the MP phase is entropically favored than the DS phase for $u \! < \! u_{\rm mic}$, starting from an initial configuration residing in the DS phase, the system then is destined to arrive at the MP phase and then reach equilibrium within this phase. The MP phase is stable and meaningful only in the microcanonical ensemble of fixed energy density. Equilibrium MP configurations contain extensive information about the planted ground state as they are already quite similar to it. The ferromagnetic ground state is easily reachable from this intermediate microcanonical phase by gradually decreasing the energy of the system, or it may be inferred based on the statistical properties of the MP configurations. The MSSB route therefore serves as a conceptually straightforward mechanism to circumvent all the low-energy spin glass traps.

The MSSB transition between the DS and MP phases is a discontinuous process.  This indicates that the DS-to-MP transition should be an extremely rare event in microcanonical simulations. We find that this process can be accelerated by the restart strategy of microcanonical random walk.  We also compare the efficiency of this microcanonical fixed-energy route with that of the conventional canonical fixed-temperature Monte Carlo. We prove that if the two routes start from the same initial state (point A of Fig.~\ref{fig:TwoRoutes}), the canonical route is more efficient than the microcanonical one; while if they pass through the same intermediate state (point B of Fig.~\ref{fig:TwoRoutes}), the microcanonical route is more efficient than the canonical one. As a byproduct concerning the conventional SA algorithms, our work indicates that as the mean energy density of the sampled configurations is approaching the critical value $u_{\rm d}$ of spin glass phase transition, the inverse temperature $\beta$ of the SA process should stop to be further elevated.

We also suggest the potential possibility of solving the planted $p$-spin model by machine-learning methods. We sample a large number of independent microcanonical configurations of the DS phase for the the most studied systems of $p \! = \! 3$ (each interaction involving three vertices).Very interestingly, we find that these DS configurations contain information about the unique planted ground state. It may then be possible to infer the planted solution from the sampled DS configurations through machine-learning techniques. We thus construct an unsupervised learning problem of inferring the planted ground state from the data generated by microcanonical samplings.  In the future, we will try to solve this inference problem using machine-learning algorithms, and will also extend this work to other hard planted systems.

\section{Theoretical predictions}

Consider a planted $p$-spin interaction system in which each vertex $j \! \in\! \{1, \ldots, N\}$ participates in $K$ clauses (interactions) and each clause involves $p$ randomly chosen vertices. For simplicity and to be concrete, we set $p\! = \! 3$ in all the following numerical computations. The total number of clauses is $M \! = \! K N / p$. The energy of a generic configuration $\vec{\sigma}\! \equiv\! (\sigma_1, \ldots, \sigma_N)$ is
\begin{equation}
  \label{eq:Esigma}
  E(\vec{\sigma}) = \sum\limits_{a=1}^{M} \Bigl[ - J_a \prod\limits_{j \in \partial a} \sigma_j \Bigr] \; ,
\end{equation}
where $\sigma_j \! \in \! \pm 1$ is the spin of vertex $j$ and $\partial a$ denotes the set of vertices constrained by clause $a$. We denote by $u$ the energy density of the system, namely
\begin{equation}
  u  =  \frac{ E(\vec{\sigma})}{N}  \; .
\end{equation}

There is a planted spin configuration $\vec{\sigma}^0\! \equiv\! (\sigma_1^0, \ldots, \sigma_N^0)$ dictating the coupling constant of clause $a$, such that
\begin{equation}
  J_a  = \left\{
  \begin{array}{ll}
   + \prod\limits_{j\in \partial a} \sigma_j^0 & \quad \quad (\textrm{probability}
    \  1  - \varepsilon) \; , \\
    & \\
    -\prod\limits_{j\in \partial a} \sigma_j^0 & \quad \quad (\textrm{probability} \  \varepsilon) \; .
  \end{array}
  \right.
\end{equation}
The parameter $\varepsilon$ is the noise level of the above planting rule. When $\varepsilon \! > \! 0$ model (\ref{eq:Esigma}) can then be interpreted as the Sourlas code of $p$-body interactions~\cite{Sourlas-1989,Huang-Zhou-2009}. When $\varepsilon$ is sufficiently small the planted configuration $\vec{\sigma}^0$ is a ground state of Eq.~(\ref{eq:Esigma}) and it may lie extensively below all the other minimum-energy configurations (Fig.~\ref{fig:Landscape}). We define the overlap (magnetization) $m$ of configuration $\vec{\sigma}$ with respect to $\vec{\sigma}^0$ as
\begin{equation}
  \label{eq:mexp}
  m = \frac{1}{N} \sum\limits_{j = 1}^{N} \sigma_j \sigma_j^0 \; .
\end{equation}

We introduce an inverse temperature $\beta$ as the conjugate of the energy $E(\vec{\sigma})$. The partition function $Z$ of the system is then defined as
\begin{equation}
  Z = \sum\limits_{\vec{\sigma}} e^{-\beta E(\vec{\sigma}) }
  = \sum\limits_{\vec{\sigma}} \prod\limits_{a=1}^{M} \exp\Bigl( \beta J_a
  \prod\limits_{j\in \partial a} \sigma_j \Bigr) \; .
\end{equation}
The equilibrium Boltzmann distribution of observing a particular configuration $\vec{\sigma}$ is then
\begin{equation}
  \label{eq:Boltz}
  P_{\textrm{eq}}( \vec{\sigma} ) = \frac{1}{Z(\beta)} e^{-\beta E(\vec{\sigma})}
  \; .
\end{equation}
In the canonical statistical ensemble the system exchanges energy with the environment. Then $\beta$ is the inverse temperature of the environment, and the mean energy density $u$ of the system is determined by $\beta$. The free energy density $f$ of the system is defined as
\begin{equation}
  f  = -\frac{1}{N} \frac{ \ln Z }{ \beta} \; .
\end{equation}
And the entropy density $s$ is computed as
\begin{equation}
  \label{eq:sval}
  s = \beta \bigl( u - f \bigr) \; .
\end{equation}
\begin{figure}
  \centering
  \subfigure[]{
    \label{fig:HRRK10L3:f}
    \includegraphics[angle=270, width=0.469\linewidth]{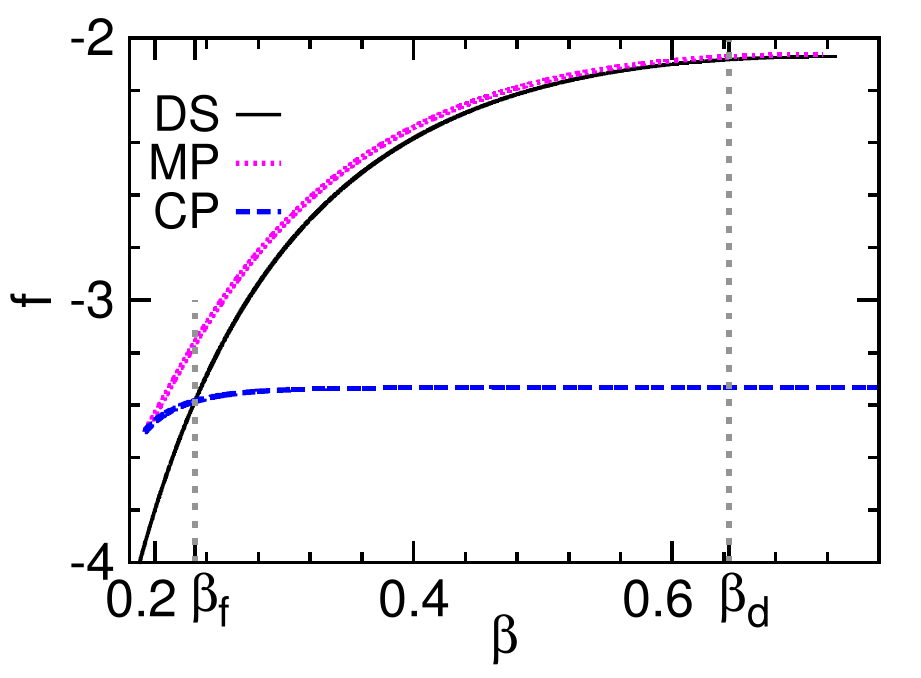}
  }
  \subfigure[]{
    \label{fig:HRRK10L3:s}
    \includegraphics[angle=270, width=0.469\linewidth]{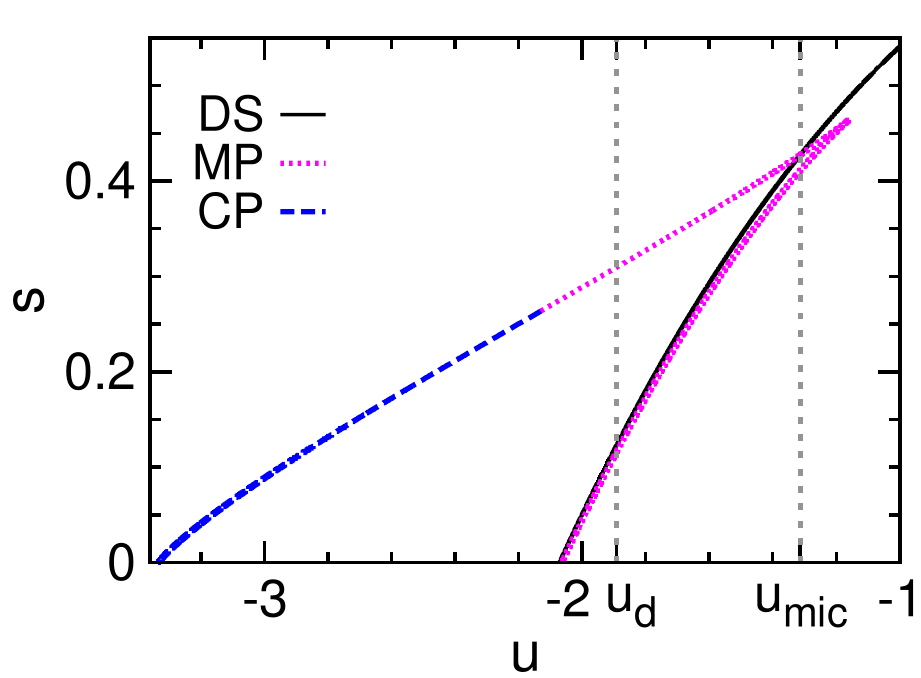}
  }
  
  \subfigure[]{
    \label{fig:HRRK10L3:b}
    \includegraphics[angle=270, width=0.469\linewidth]{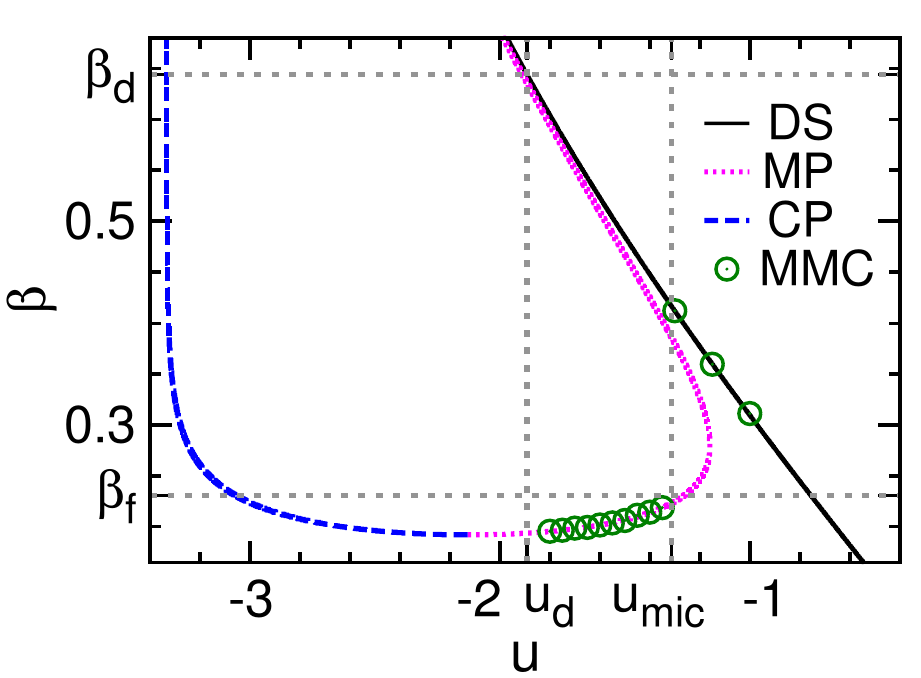}
  }
  \subfigure[]{
    \label{fig:HRRK10L3:m}
    \includegraphics[angle=270, width=0.469\linewidth]{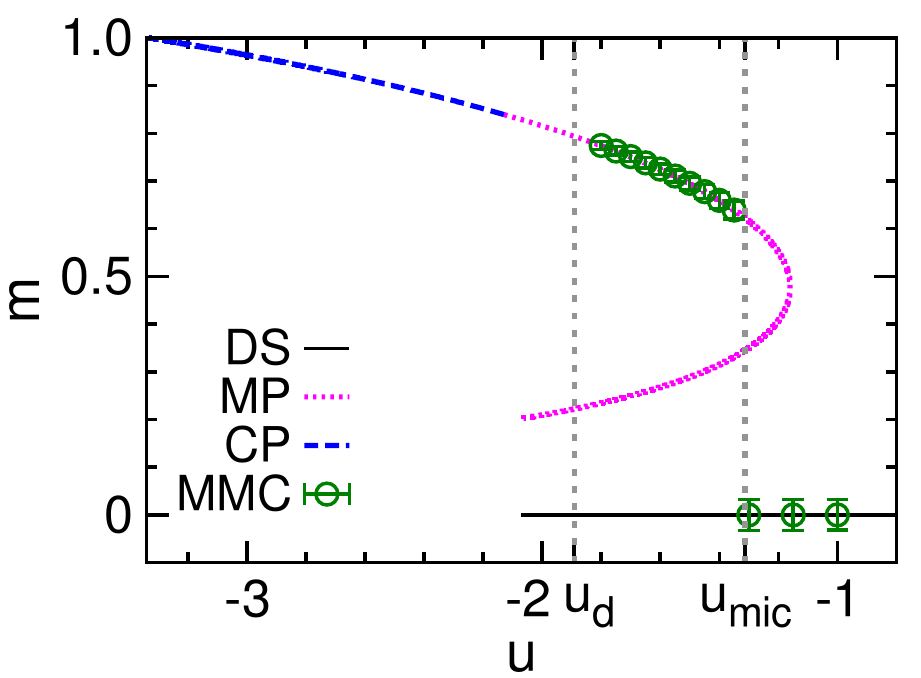}
  }
  \caption{\label{fig:HRRK10L3}
    The DS (solid line), MP (dotted line), and CP (long dashed line) fixed points of the mean-field theory for the planted $3$-body model on random graphs of degree $K\! =\! 10$ at noise $\varepsilon \! = \! 0$. (a) Free energy density $f$ versus canonical inverse temperature $\beta$. (b) Entropy density $s$ versus energy density $u$. (c) Microcanonical inverse temperature $\beta$ versus $u$. (d) Mean overlap $m$ versus $u$. $\beta_{\rm d}$ and $u_{\rm d}$: the critical inverse temperature and energy density at the dynamical SG phase transition; $u_{\rm mic}$: the critical energy density at the MSSB phase transition; $\beta_{\rm f}$: the critical inverse temperature at the canonical ferromagnetic phase transition. The predicted MSSB phase transition is confirmed by MMC simulations (circles) on a single problem instance of size $N \! = \! 960$.
  }
\end{figure}

In the microcanonical statistical ensemble, the system is isolated from the environment and its energy density $u$ is a fixed value, then $\beta$ is interpreted as the microcanonical (intrinsic) inverse temperature of the system. With respect to the Boltzmann distribution (\ref{eq:Boltz}), $\beta$ in the microcanonical statistical ensemble serves as a Lagrange multiplier such that the mean energy of the sampled configurations according to Eq.~(\ref{eq:Boltz}) is equal to a prescribed value $u$. In other words, $\beta$ is determined by $u$ in the microcanonical statistical ensemble.

This random-graph system has been intensively studied by the mean-field cavity method of statistical mechanics~\cite{Mezard-Parisi-2001,Mezard-Montanari-2006,Matsuda-etal-2011,Zhou-2015}. For the ensemble of regular random graph of vertex degree $K$, the mean-field equations for the mean overlap $m$, the mean energy density $u$, the free energy density $f$, and the entropy density $s$ are easy to solve. The detailed equations are listed in Appendix~\ref{app:mfr}.

The mean-field results obtained at $\varepsilon\! = \! 0$ and $K \! = \! 10$ are demonstrated in Fig.~\ref{fig:HRRK10L3}, and the qualitatively identical results obtained for $K\! = \! 4$ are shown in Fig.~\ref{fig:HRRK4L3} of Appendix~\ref{app:mfr}. Some aspects of these theoretical results are similar to what were found for the Potts model~\cite{Zhou-2019}, but there is a key qualitative difference to be discussed at the end of this section. First, let us interpret these theoretical predictions from the perspective of microcanonical thermodynamics~\cite{Gross-2001}.

The configuration space has a disordered symmetric phase whose mean magnetization is zero in the thermodynamic limit. This DS phase contains the paramagnetic spin configurations and it corresponds to the paramagnetic phase of the canonical ensemble. As the energy density $u$ decreases to the spin glass (SG) dynamical transition point $u_{\rm d}$ this DS phase suddenly divides into an exponential number of ergodicity-broken macro-states, and the system then enters into the SG phase~\cite{Franz-etal-2001}. The value of $u_{\rm d}$ is independent of $\varepsilon$ because of gauge symmetry~\cite{Matsuda-etal-2011} and can be determined by the tree-reconstruction method \cite{Mezard-Montanari-2006} (see also textbooks \cite{Mezard-Montanari-2009,Zhou-2015}).

Because of planting, a stable canonical polarized (CP) phase containing configurations similar to $\vec{\sigma}^0$ exist in the configuration space at sufficiently low energies. This CP phase is simply the conventional ferromagnetic phase and its overlap $m$ is close to unity.

The microcanonical polarized phase, corresponding to the unstable fixed point of the mean field theory with higher free energy density [Fig.~\ref{fig:HRRK10L3:f}], serves as the watershed between the DS and CP phases [Fig.~\ref{fig:HRRK10L3:b}]. Individual configurations of the MP phase have positive overlaps $m$ and therefore are similar to $\vec{\sigma}^0$ and to configurations of the CP phase, but they are unstable in the canonical ensemble due to entropic effect.

The entropy density $s(u)$ of the MP phase as a function of energy density $u$ has two branches [Fig.~\ref{fig:HRRK10L3:s}]. The higher-entropy branch corresponds to the MP configurations that are stable at fixed $u$, and its $s(u)$ is convex. The MP phase starts to surpass the DS phase in entropy density at the critical energy density $u_{\rm mic}$, leading to a kink of $s$ and indicating a discontinuous microcanonical spontaneous symmetry-breaking phase transition~\cite{Zhou-2019}. This phase transition is characterized by a jump in the mean overlap $m$ and a drop in the microcanonical inverse temperature $\beta$, which are verified by our simulation results [Fig.~\ref{fig:HRRK10L3:b} and \ref{fig:HRRK10L3:m}]. The lower-entropy MP branch, on the other hand, always has lower entropy than that of the DS phase and its $s(u)$ is concave; it marks the watershed between the MP and DS phases in the microcanonical ensemble.

We observe that the MSSB transition energy density $u_{\rm mic}$ is located above the SG transition point $u_{\rm d}$ when the noise level $\varepsilon$ is low (Table~\ref{tab:RRresults}), for all values of degree $K \! \geq \! 4$. This means that there is an energy density range $u \in (u_{\rm d}, u_{\rm mic})$ within which the MP configurations are exponentially dominating over the DS configurations. In an infinitely slow microcanonical annealing process, the SG phase will then not be encountered, instead the system will jump from the DS phase to the MP phase at the higher energy density point $u_{\rm mic}$.  In principle,  it is then possible to reach the MP phase by inducing a MSSB transition at fixed value of $u \in \! ( u_{\rm d}, u_{\rm mic})$, avoiding the spin glass traps of the canonical ensemble. We will continue to discuss the practical feasibility of this proposal in the next section.

We find that the DS fixed point of the $p$-spin model is always locally stable, down to the minimal energy density at which the entropy density becomes zero [Fig.~\ref{fig:HRRK10L3:s}]. Consequently, a gap of the overlap $m$ always exists between the DS and MP phases down to this minimal energy density [Fig.~\ref{fig:HRRK10L3:m}]. This feature is significantly different from what was observed in the Potts model (see Fig.~2(d) of Ref.~\cite{Zhou-2019}). The DS phase of the Potts model becomes unstable below certain threshold energy density, and by fixing the energy density below this threshold value, the system will then evolve from the DS phase to the MP phase gradually. Such a gradual evolution will not be possible for the $p$-spin model because of the positive gap in the order parameter $m$.

\begin{table}
  \caption{
    \label{tab:RRresults}
    Overlap jump $\Delta m$, microcanonical inverse temperature drop $\Delta \beta$, critical energy density $u_{\rm mic}$,  at the MSSB phase transition of the planted $3$-body model on random graphs of degree $K$ (noise $\varepsilon\! =\! 0$). The critical energy densities $u_{\rm d}$ are also listed (based on Table 4.2 of \cite{Zhou-2015}).
  }
  \centering
  \vspace*{0.2cm}
  \begin{tabular}{l|rrrrrrr}
    \hline \hline
    $K$          & $4$ & $5$ & $6$ & $7$ & $8$ & $9$ & $10$ \\ \hline
     $\Delta m$  & $0.906$ & $0.821$ & $0.758$ & $0.712$ & $0.676$ & $0.647$ & $0.623$ \\
    $\Delta \beta$  & $-0.595$ & $-0.409$ & $-0.324$ & $-0.274$ & $-0.239$ & $-0.214$ & $-0.193$ \\
    $u_{\rm mic}$ & $-1.115$ & $-1.167$ & $-1.204$ & $-1.235$  & $-1.264$ & $-1.289$ & $-1.313$ \\
    $u_{\rm d}$   &  $-1.159$ & $-1.316$ & $-1.453$ & $-1.575$ & $-1.687$  & $-1.792$ & $-1.892$ \\ \hline \hline
  \end{tabular}
\end{table}

We notice that the special case of degree $K \! = \! 3$ is qualitatively different, for which the DS phase always dominates over the MP and CP phases at all the energy density values (see Fig.~\ref{fig:HRRK3L3} of Appendix~\ref{app:mfr}). An equilibrium MSSB phase transition is then absent for this special system.

\section{Numerical simulation results}

In this section, we discuss the numerical results obtained by two different microcanonical Monte Carlo (MMC) algorithms, and we also compare the efficiency of the MMC dynamics with that of the conventional canonical Monte Carlo (CMC) dynamics. 

\subsection{The single-flip MMC algorithm}

We first implement a simple MMC algorithm to explore configurations at the vicinity of an objective energy $E_{\rm o} \! \equiv \! N u$, whose energy density $u\! \in \! (u_{\rm d}, u_{\rm mic})$. An initial configuration $\vec{\sigma}$ with energy $E(\vec{\sigma}) \! \leq \! E_{\rm o}$ can be sampled according to the equilibrium Boltzmann distribution (\ref{eq:Boltz}) at a suitably chosen inverse temperature $\beta$. Starting from this initial configuration, the following single-flip trial is performed at each elementary time interval $1/N$ of the MMC evolution: (1) a vertex $j$ is picked uniformly at random from the $N$ vertices and the flip $\sigma_j \! \rightarrow \! -\sigma_j$ is proposed; (2) if the new configuration energy does \emph{not} exceed $E_{\rm o}$, this flip is accepted, otherwise it is rejected and the old spin $\sigma_j$ is kept for vertex $j$. One sweep (unit time) of this MMC process corresponds to $N$ consecutive single-flip trials.

This single-flip dynamics obeys detailed balance since the forward and backward transitions between two configurations are equally likely~\cite{Creutz-1983,Rose-Machta-2019}. This microcanonical detailed balance condition ensures that all the visited configurations along the evolution trajectory share the same statistical weight. We sample configurations at fixed time intervals and record their overlap values $m$ and energies $E$. The microcanonical inverse temperature of the system at energy density $u$ is estimated through
\begin{equation}
  \beta = \frac{1}{4} \ln \Bigl( 1 + \frac{4}{E_{\rm o} -
    \bigl\langle E(\vec{\sigma}) \bigr\rangle }
  \Bigr) \; ,
\end{equation}
where $\bigl\langle E(\vec{\sigma}) \bigr\rangle$ is the averaged energy of the sampled configurations~\cite{Creutz-1983}.

\begin{figure}
  \centering
  \subfigure[]{
    \label{fig:WTimeRRL3a}
    \includegraphics[angle=270,width=0.469\linewidth]{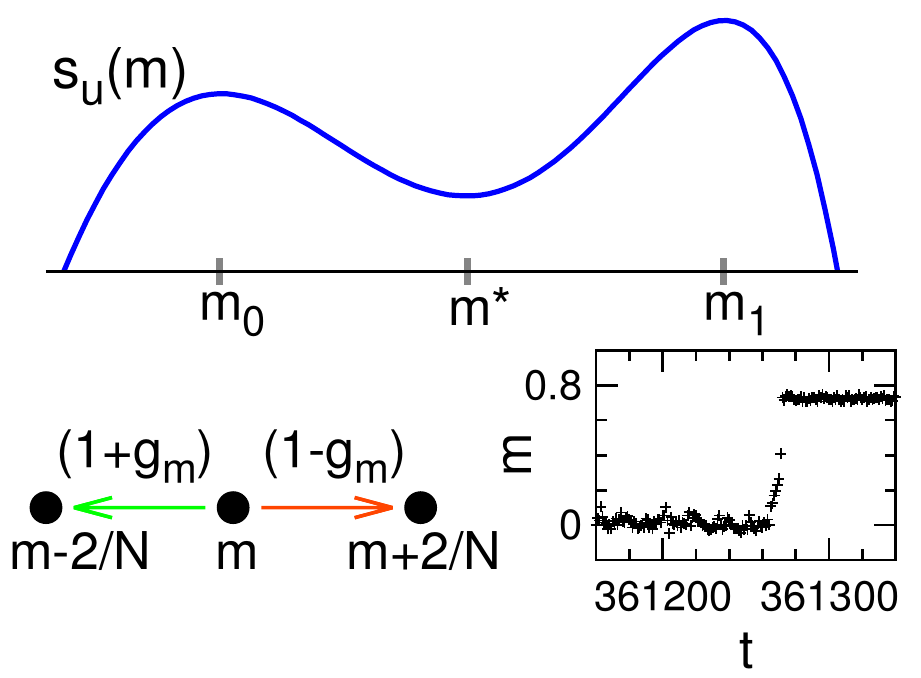}
  }
  \subfigure[]{
    \label{fig:WTimeRRL3b}
    \includegraphics[angle=270,width=0.469\linewidth]{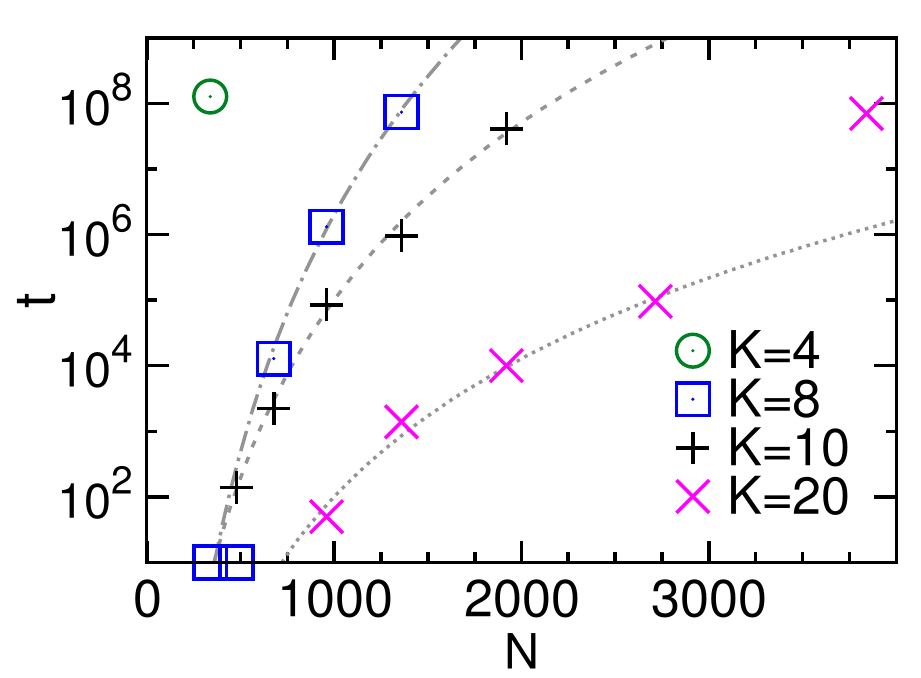}
  }
  \caption{
    \label{fig:WTime}
    An entropy barrier has to be crossed to achieve a MSSB transition. (a) A schematic curve of the entropy density $s_u(m)$ of the planted $3$-body model at fixed energy density $u$ and noise $\varepsilon \! = \! 0$, showing a local maximum at overlap $m_0\! \approx \! 0$, a global maximum at $m_1$, and a minimum at $m^*$. The MMC dynamics is equivalent to a one-dimensional biased random walk with bias ratio $g_m$.  The example evolution trajectory shows a MSSB event at $u\! = \! -1.61$ in a random graph of degree $K\! = \! 10$ and size $N \! = \! 960$ (one unit of MMC time $t$ means $N$ single-flip trials). (b) The minimal number of sweeps $t$ needed for a MSSB transition to occur, with each point being the length of the shortest trajectory among $24$ independent single-flip MMC runs on a single network: $K \! = \! 4$, $u \! = \! -1.133$ (circle); $K \! = \! 8$, $u \! = \! -1.6$ (squares); $K \! = \! 10$, $u \! = \! - 1.61$ (pluses); $K \! = \! 20$, $u \! = \! - 2.2$ (crosses). Curves are power-law ($\propto\! N^b$) relations with $b \! = \! 7, 9, 12$ from bottom to top.
  }
\end{figure}

Our MMC dynamics is an unbiased random walk within the microcanonical configuration space. The evolution trajectory is expected to be ergodic since the energy density $u$ is above the dynamical spin glass transition point $u_{\rm d}$. As evolution time goes to infinity every configuration of this space should have equal frequency to be visited, and the system will then certainly be in the MP phase since this phase is exponentially dominant in statistical weight. Empirically, we have observed that this MMC dynamics does indeed achieve MSSB in small-sized systems [Fig.~\ref{fig:HRRK10L3:b} and Fig.~\ref{fig:WTimeRRL3a}], while the evolution time needed to witness such a transition increases quickly with system size $N$ [Fig.~\ref{fig:WTimeRRL3b}].  It is easier for the MSSB transition to occur in networks of larger degrees $K$. This is consistent with the mean-field theoretical results, which reveal that the entropy barrier and the gap of the order parameter $m$ both decreases with $K$ (Table~\ref{tab:RRresults}). The hardest RR problem instances are those with degree $K \! = \! 4$.

\begin{figure*}
  \centering
  \subfigure[]{
    \label{fig:TwoRoutes}
     \includegraphics[angle=270,width=0.26\linewidth]{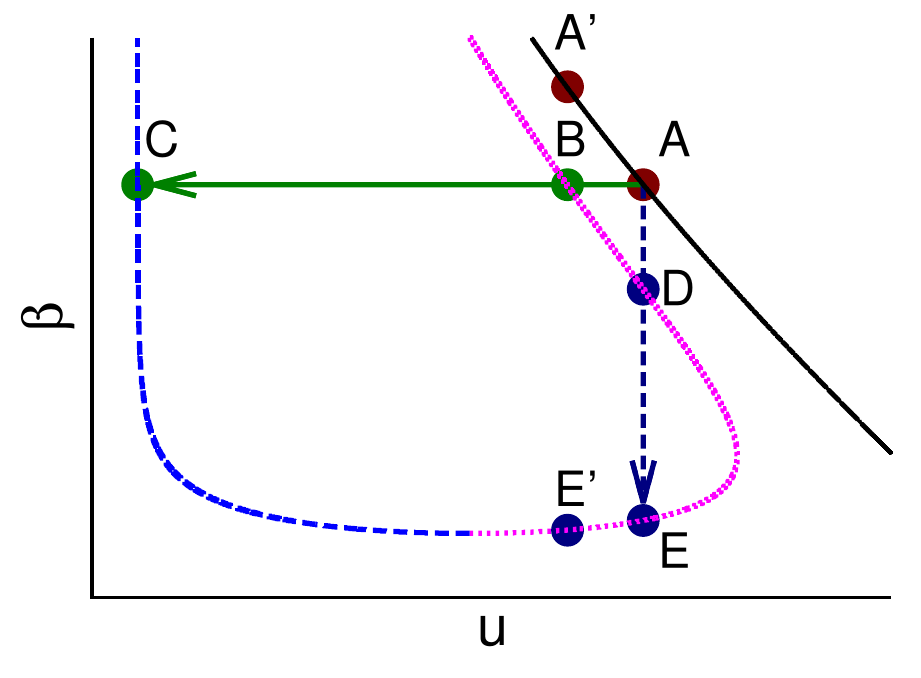}
  }
  \subfigure[]{
    \label{fig:RRK10L3gapU}
     \includegraphics[angle=270, width=0.26\linewidth]{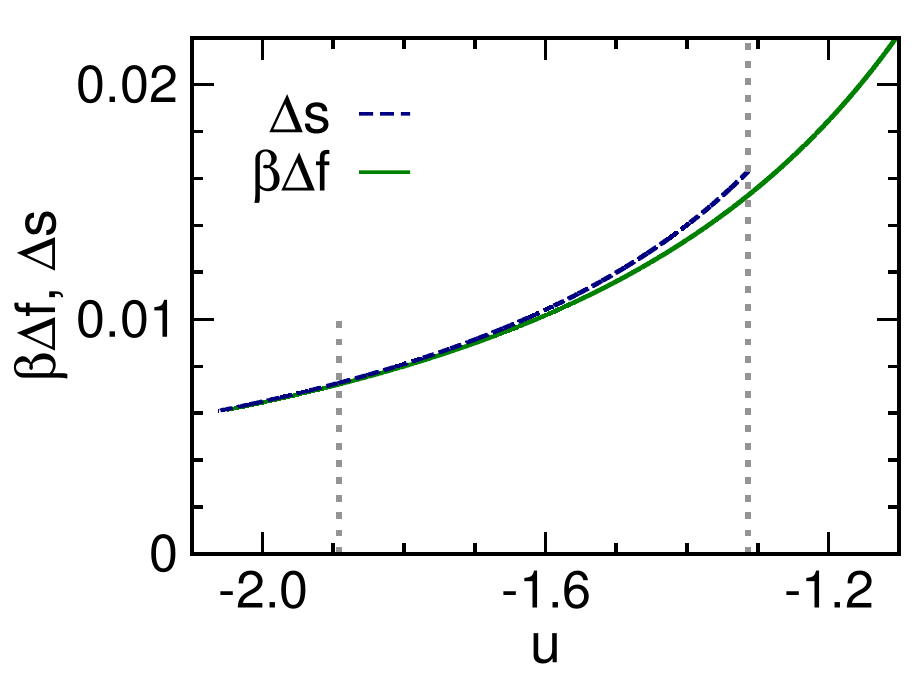}
  }
  \subfigure[]{
    \label{fig:WTN1008}
    \includegraphics[angle=270,width=0.26\linewidth]{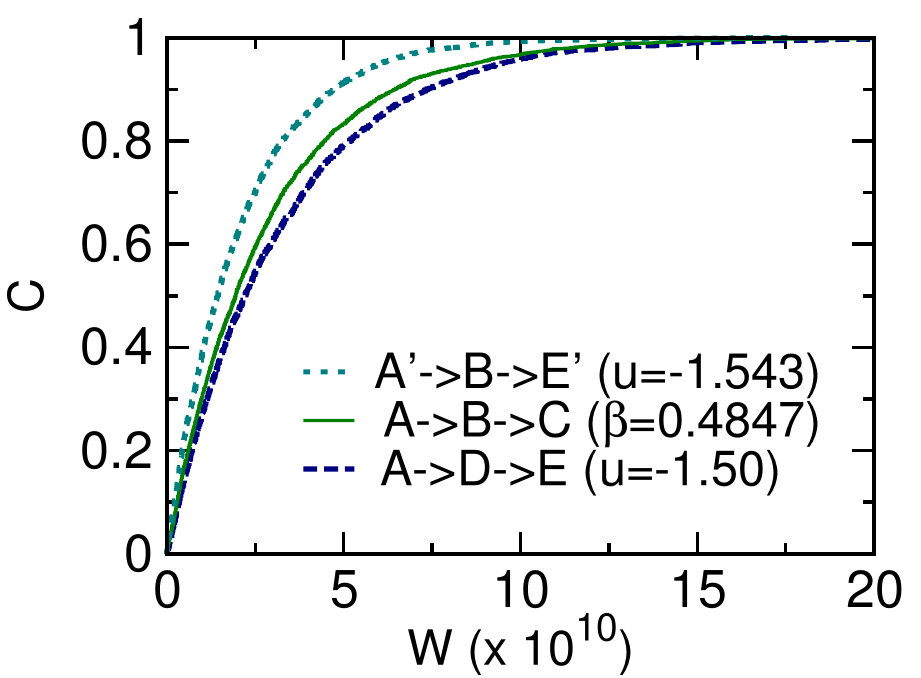}
  }
  \caption{
     \label{fig:efficiency}
    Two different routes to escape the disordered symmetric phase. (a) Schematic plot showing the canonical route ($A\! \rightarrow \! B \! \rightarrow \! C$) at fixed inverse temperature $\beta$ and the microcanonical route ($A\! \rightarrow \! D\! \rightarrow \! E$) at fixed energy density $u$. The black solid line represents the DS phase (containing points $A$ and $A^\prime$); the blue long-dashed line is the CP phase (containing point $C$); the red dotted line is the MP phase (containing points $B, D, E$, and $E^\prime$). (b) The rescaled free energy barrier $\beta \Delta f$ of the canonical route (solid line) and the entropy barrier $\Delta s$ of the microcanonical route (dashed line), as a function of the energy density $u$ of the initial configuration. These are theoretical results obtained at $K \! = \! 10$ and noise $\varepsilon \! = \! 0$. The vertical dotted lines mark $u_{\rm d}$ (left) and $u_{\rm mic}$ (right). (c) Cumulative distribution $C(W)$ on the total number of spin flips $W$ performed to escape the DS phase, obtained by simulating $4000$ independent hybrid-flip CMC evolution trajectories at $\beta \! = \! 0.4847$ (route $A\! \rightarrow\! B \! \rightarrow\! C$, solid line), $4000$ independent hybrid-flip MMC evolution trajectories at $u \! = \! -1.50$ (route $A\! \rightarrow \! D \! \rightarrow \! E$, dashed line) and another $4000$ independent trajectories at $u \! = \! -1.543$ (route $A^\prime \! \rightarrow \! B \! \rightarrow \! E^\prime$, dotted line) on a single graph instance of size $N\! = \! 1008$ and degree $K \! = \! 10$ (noise $\varepsilon \! = \! 0$). 
  }
\end{figure*}

The MSSB transition is an entropy-barrier crossing process [Fig.~\ref{fig:WTimeRRL3a}, and the route $A\! \rightarrow \! D\! \rightarrow \! E$ in Fig.~\ref{fig:TwoRoutes}]. We can regard the MMC dynamics on the coarse-grained level of overlap $m$ as a one-dimensional random walk under a potential field $-s_u(m)$, where $s_u(m)$ is the entropy density of configurations at fixed energy density $u$ having overlap $m$ with the planted configuration $\vec{\sigma}^0$. At each elementary trial, $m$ may change to $m \pm 2/N$ with probability proportional to $(1 \mp g_m)/2$. Here the bias ratio is
\begin{equation}
  g_m  =  \frac{1 - e^{2 s_u^\prime(m)}}{1 + e^{2 s_u^\prime(m)}} \; ,
\end{equation}
with $s_u^\prime (m ) \! \equiv \! {\rm d} s_u(m)/{\rm d} m$ being the slope of $s_u(m)$ at $m$. Because $s^\prime(m) < 0$ in $m\! \in \! (0, m^*)$ with $m^*$ being the watershed point of $s_u(m)$ [Fig.~\ref{fig:WTimeRRL3a}, and point $D$ of Fig.~\ref{fig:TwoRoutes}], the occurrence of MSSB is an exponentially rare first-passage event characterized by a waiting time of exponential order $O(e^{N \Delta s})$, where $\Delta s \! = \! s_u(0) - s_u(m^*)$ is the barrier height~\cite{Weiss-1981,Khantha-Balakrishnan-1983}. Consequently, this simple random walk strategy is practical only for systems of sizes
\begin{equation}
  N <  \frac{20}{\Delta s} \; .
\end{equation}

The entropy barrier $\Delta s$ decreases with energy density $u$ [Fig.~\ref{fig:RRK10L3gapU}]. Therefore, we can decrease the entropy barrier $\Delta s$ by fixing $u$ of the MMC dynamics to a lower value. This will help reducing the waiting time of the MSSB transition. On the other hand, at lower energy density values the number of transition paths from point $A$ to point $D$ of Fig.~\ref{fig:TwoRoutes} becomes much more rarer and then disappears at $u \! \approx \! u_{\rm d}$. This later kinetic effect will lead to an increase in the waiting time. There should be an optimal energy density $u$, at which the MSSB transition is easiest to observe. We leave this interesting issue to future studies.

As a minor point, we notice that the watershed entropy density $s_u(m^*)$ of Fig.~\ref{fig:WTimeRRL3a} decreases as the energy density $u$ decreases and it will reach zero at certain critical energy density value (denoted as $u_{\rm meb}$) which marks the point of strong microcanonical ergodicity-breaking. When $u \! <  \! u_{\rm meb}$ there will be an extended region of the intermediate overlap values $m$ within which there is no any configuration of energy density $u$. In other words, it is impossible to reach from a disordered initial configuration to a highly ordered final configuration through a sequence of small changes of the configuration  at fixed energy density $u \! < \! u_{\rm meb}$. We find that $u_{\rm med}$ is below $u_{\rm d}$ for all the degrees $K \! \geq \! 3$ (see Figs.~\ref{fig:HRRK10L3:s}, \ref{fig:HRRK4L3:s} and \ref{fig:HRRK3L3:s}), so this strong microcanonical ergodicity-breaking is not really relevant for our MSSB mechanism which occurs at $u \! > \! u_{\rm d}$.

\subsection{The hybrid-flip MMC and CMC algorithms}

Besides the microcanonical route $A\! \rightarrow \! D \! \rightarrow \! E$ as depicted in Fig.~\ref{fig:TwoRoutes}, there is also the conventional canonical route ($A \! \rightarrow \! B \! \rightarrow \! C$) to escape the disordered symmetric phase. Will this canonical route be blocked by the low-energy disordered spin glass states? If the planted ground state could be successfully approached following either the microcanonical route or the canonical route, which of these two routes will be more efficient? We explore these two questions in this and the next subsections.

To be fair in comparing the MMC and CMC routes, we design a hybrid-flip Monte Carlo dynamics which is rejection-free both at fixed energy density $u$ (MMC) and at fixed inverse temperature $\beta$ (CMC). Besides the action of flipping a single spin, the hybrid-flip dynamics differs from the single-flip dynamics by allowing the action of flipping a pair of spins in one elementary updating step.

An elementary step of the hybrid-flip MMC dynamics goes as follows: (1) a vertex $j$ is picked uniformly at random from the $N$ vertices and its spin $\sigma_j$ is flipped, causing a change $\Delta E$ to the energy; (2) if and only if the energy of the new configuration exceeds the objective value $E_{\rm o}$, an additional flip is performed on the spin $\sigma_k$ of a vertex $k$ to bring the energy back to its original value ($k$ is picked uniformly at random from all the vertices capable of causing a change $-\Delta E$ to the energy). This microcanonical hybrid-flip rule obeys microcanonical detailed balance, so all the visited configurations on the MMC evolution trajectory have the same statistical weight.

An elementary step of the hybrid-flip CMC dynamics is quite similar: (1) a vertex $j$ is picked uniformly at random from the $N$ vertices and its spin $\sigma_j$ is flipped, causing a change $\Delta E$ to the energy; (2) if $\Delta E$ is positive, then with probability $1\! -\! e^{-\beta \Delta E}$ an additional flip is performed on the spin $\sigma_k$ of a vertex $k$ to bring the energy back to its original value ($k$ is picked uniformly at random from all the vertices capable of causing a change $-\Delta E$ to the energy). This canonical hybrid-flip rule obeys canonical detailed balance and the visited configurations on the CMC evolution trajectory are governed by the Boltzmann distribution (\ref{eq:Boltz}).

We carry out hybrid-flip MMC and hybrid-flip CMC simulations on a single random graph instance of degree $K \! = \! 10$ and size $N\! = \! 1008$. A total number of $4000$ independent MMC evolution trajectories and $4000$ independent CMC evolution trajectories are sampled with the help of the TRNG pseudo-random number library~\cite{Bauke-Mertens-2007}, all of these trajectories start from the same initial equilibrium configuration which is located in the DS phase (inverse temperature $\beta \! = \! 0.4847$ and energy density $u\! = \! -1.50$). We regard the DS phase as successfully escaped when the overlap $m$ with the planted configuration exceeds $0.6$  for the first time, and the evolution is then terminated and the total number $W$ of spin flips along the whole evolution trajectory is returned. We refer to $W$ as the effort to escape DS. Figure~\ref{fig:WTN1008} shows the cumulative distribution profile $C(W)$ of the effort $W$ over the $4000$ samples, obtained at fixed inverse temperature $\beta$ (CMC) or fixed energy density $u$ (MMC).

We find that the effort (number of spin flips) $W$ fluctuates over a wide range from $10^6$ up to $10^{12}$ among repeated runs (all of which start from the same initial configuration). This indicates that restart might be an optimal strategy if the evolution trajectory fails to escape the DS phase within a certain threshold value of $W$.  The cumulative distribution function $C(W)$ is concave in shape, both for the MMC dynamics and the CMC dynamics. For example, within an effort $W_1\! \approx \! 6.071\times 10^9$ the success probability of escaping the DS phase is $C(W_1) \! = \! 20\%$ by the CMC dynamics, while this success probability only increases to $C(W_2) \! = \! 35.6\%$ when the effort is doubled to $W_2\! = \! 1.214\times 10^{10}$ [Fig.~\ref{fig:WTN1008}]. This concave property of $C(W)$ guarantees that restart is an optimal strategy. For example, we could restart the CMC or the MMC evolution from the fixed initial configuration if the effort of the current trajectory exceeds $W_1$, and the resulting cumulative distribution of the effort will change to be approximately linear. 

\subsection{Entropy barrier and free energy barrier}

Our CMC simulation results clearly demonstrate that the low-energy spin glass states are not encountered by the canonical route $A\! \rightarrow \! B \! \rightarrow \! C$. This may be easy to understand: the canonical route occurs at a much lower inverse temperature than the critical value $\beta_{\rm d}$ where the spin glass dynamical phase transition occurs, and the energy density of the watershed point $B$ separating the DS and the CP phases is considerably higher than the critical value $u_{\rm d}$ of spin glass dynamical phase transition.

Figure~\ref{fig:WTN1008} demonstrates that, when starting from the same initial configuration, the CMC dynamics is more efficient than the MMC dynamics. For example, the MMC dynamics needs a larger effort $W_1 \! \approx \! 7.093 \times 10^9$ to achieve a success probability $C(W_1) \! = \! 20\%$. The empirical observation of CMC being more efficient than MMC is consistent with the theoretical predictions on the entropy-barrier height $\Delta s$ and the rescaled free energy barrier height $\beta \Delta f$ [Fig.~\ref{fig:RRK10L3gapU}]. We could prove that $\Delta s > \beta \Delta f$ as follows.

The rescaled free energy barrier ($\beta \Delta f$) of the canonical route $A\! \rightarrow \! B \! \rightarrow \! C$ is
\begin{equation}
  \label{eq:deltabF}
  \beta \Delta f =  \beta_A (f_B - f_A) \; .
\end{equation}
Here $\beta_X$, $u_X$, $f_X$, and $s_X$ denote the inverse temperature, energy density, free energy density, and entropy density at the phase point $X \in \{A, B, \ldots\}$ of Fig.~\ref{fig:TwoRoutes}. The entropy barrier ($\Delta s$) of the microcanonical route $A\! \rightarrow \! D \! \rightarrow \! E$) is
\begin{equation}
  \label{eq:dsexp}
  \Delta s = s_A - s_D \; .
\end{equation}
Becuase of the relationship (\ref{eq:sval}) and the fact that $\beta_B = \beta_A$ and $u_D = u_A$, we know that
\begin{eqnarray}
  \Delta s & = &
  \beta \Delta f + \bigl[s_B - s_D + \beta_B (u_D - u_B) \bigr]
  \nonumber \\
  & \approx &  \beta \Delta f + \frac{1}{2}
  (\beta_B - \beta_D) ( u_D - u_B)
  \nonumber \\
  & > & \beta \Delta f \; .
\end{eqnarray}
In deriving this inequality, we have used the following approximation
\begin{eqnarray}
  \hspace*{-0.250cm}
  s_D  & \approx & s_B + \left. \frac{{\rm d} s}{{\rm d} u}\right|_{u_B}
  (u_D - u_B) +
  \frac{1}{2} \left. \frac{{\rm d}^2 s}{{\rm d} u^2}\right|_{u_B}
  ( u_D - u_B)^2
  \nonumber \\
   & = & s_B +  \beta_B ( u_D - u_B) + \frac{1}{2}
  \left. \frac{{\rm d} \beta}{{\rm d} u}\right|_{u_B}
  (u_D - u_B)^2
  \nonumber \\
  & \approx & s_B + 
  \beta_B ( u_D - u_B) + \frac{1}{2} (\beta_D - \beta_B) (u_D - u_B) \; ,
\end{eqnarray}
assuming that $u_D$ is quite close to $u_B$.

On the other hand, if we compare the CMC and MMC routes passing through the same intermediate state (say point $B$ of Fig.~\ref{fig:TwoRoutes}), namely the CMC route $A\! \rightarrow\! B\! \rightarrow\! C$ and the MMC route $A^\prime \! \rightarrow B\! \rightarrow\! E^\prime$ (points $A^\prime, B, E^\prime$ have the same energy density), we find that the entropy barrier
\begin{equation}
  \label{eq:dsprmexp}
  \Delta s^\prime = s_{A^\prime} - s_B \; ,
\end{equation}
is lower than the rescaled free energy barrier $\beta_B \Delta f$ [Eq.~(\ref{eq:deltabF})].  To prove this, we notice that
\begin{eqnarray}
  \Delta s^\prime & = & \beta_B \Delta f
  + \bigl[ s_{A^\prime} - s_A + \beta_A (u_A - u_B)  \bigr]
  \nonumber \\
  & \approx &  \beta_B \Delta f  - \frac{1}{2} (\beta_{A^\prime} - \beta_A)
    (u_{A} - u_{A^\prime}) \nonumber \\
  & < &  \beta_B \Delta f\; ,
\end{eqnarray}
as $\beta_{A^\prime} \! > \! \beta_{A}$ and $u_{A} \! > \! u_{A^\prime}$. This indicates that for all the transition trajectories passing through the same intermediate state $B$, the MMC trajectories are more efficient than the CMC trajectories. Our simulation results confirm this prediction, see the dotted line and the solid line of Fig.~\ref{fig:WTN1008}.

\subsection{Simulated energy annealing (SEA)}

\begin{figure}
  \centering
  \subfigure[]{
    \label{fig:SEAbeta}
    \includegraphics[angle=270, width=0.469\linewidth]{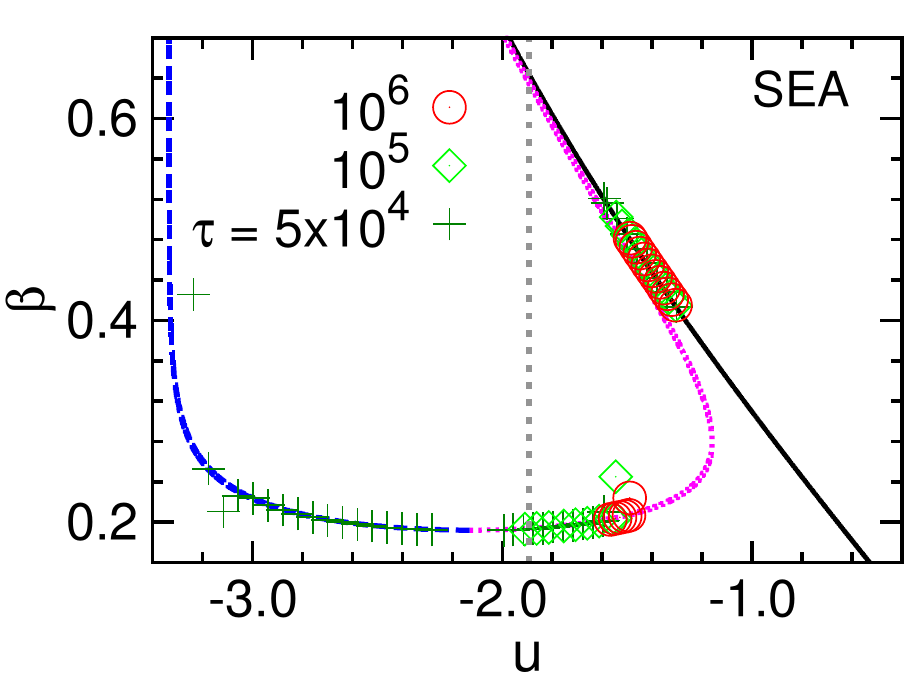}
  }
  \subfigure[]{
    \label{fig:SEAm}
    \includegraphics[angle=270, width=0.469\linewidth]{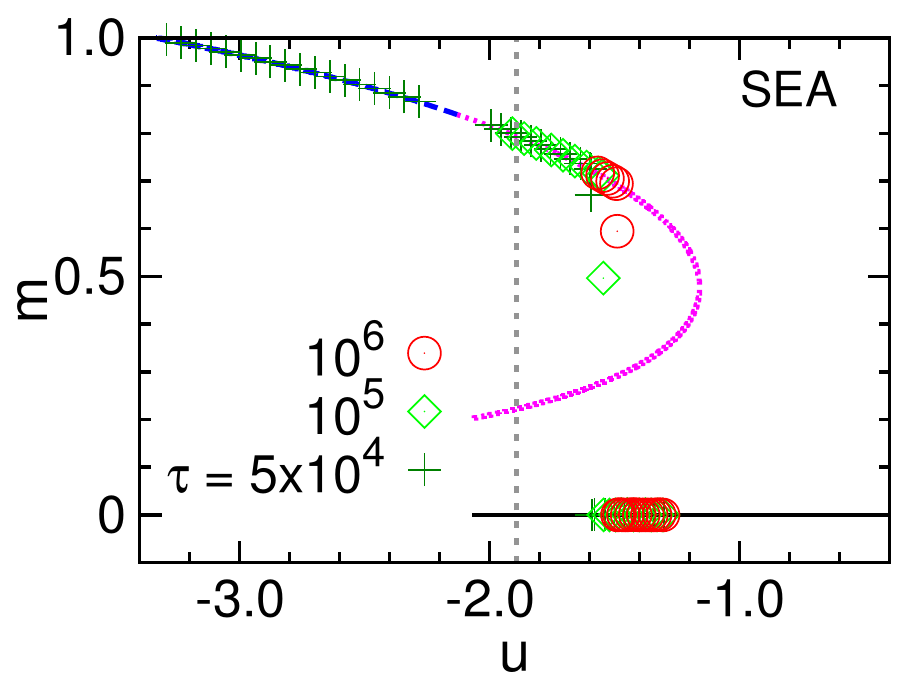}
  }
  \caption{
    \label{fig:SEAK10L3}
    Simulated energy annealing on a single regular random graph instance of size $N \! = \! 1008$ and degree $K \! = \! 10$. The energy density $u$ decreases in step of $2/N$, and $\tau$ sweeps are simulated at $u$, with $\tau\! = \! 5\times 10^4$ (pluses), $10^5$ (diamonds), or $10^6$ (circles). The microcanonical inverse temperature $\beta$ (a) and the mean overlap $m$ (b) are computed using $\tau$ configuration samples. The vertical dashed line marks the spin glass transition energy density $u_{\rm d}$.
  }
\end{figure}

We do not need to fix the energy or the inverse temperature in the Monte Carlo simulations. Instead the energy or the inverse temperature can slowly change with time. Following some previous numerical efforts~\cite{Alava-etal-2008,Ma-Zhou-2011,Rose-Machta-2019}, we implement SEA (simulated energy annealing) as follows: (1) Start from an initial configuration of high energy value $E$; (2) perform MMC simulation with at objective energy $E_{\rm o}= E$ for a total number $\tau$ of sweeps (each sweep corresponds to $N$ contiguous hybrid-flip updates); (2) decrease the objective energy by a small amount (say, $E_{\rm o}\! - \! 2 \rightarrow  E_{\rm o}$) and repeat step (2).

Some SEA evolution trajectories are shown in Fig.~\ref{fig:SEAK10L3} in the $(u, \beta)$ and $(u, m)$ phase planes. As the waiting time $\tau$ at each energy density $u$ increases the MSSB transition occurs at higher energy values $u$. This is consistent with expectations.

Similar results are also observed in the conventional simulated temperature annealing process, where the canonical inverse temperature $\beta$ gradually increases (but does not exceed $\beta_{\rm d}$). The energy density $u$ will change dramatically during this annealing process. 

In comparison with the SEA algorithm, the MMC algorithm (with restart) at a suitably chosen fixed energy density $u$ is more efficient.

\subsection{A machine-learning challenge}

Since it is exceedingly difficult to escape the DS phase by both the MMC dynamics and the CMC dynamics as the system size $N$ increases, might it be possible to retrieve the planted ground state without actually leaving the DS phase? Here we report a preliminary exploration of this interesting question.

Given a $p$-spin problem instance, it is straightforward to sample a large set of independent equilibrium configurations of the DS phase at a given energy density $u\! \in\! (u_{\rm d}, u_{\rm mic})$ by the hybrid-flip MMC algorithm. The overlaps $m$ of these configurations are of order $1 / \sqrt{N}$, so a more convenient quantity is the rescaled overlap $\lambda \equiv m \sqrt{N}$, namely
\begin{equation}
  \lambda = \frac{1}{\sqrt{N}} \sum\limits_{i=1}^{N} \sigma_i \sigma_i^0
  \; .
\end{equation}
An empirical probability distribution $P(\lambda)$ could be constructed as
\begin{equation}
  P(\lambda) = \frac{1}{\mathcal{N}}
  \sum\limits_{\ell = 1}^{\mathcal{N}}
  \delta\biggl( \lambda -
  \frac{1}{\sqrt{N}} \sum\limits_{i=1}^{N} \sigma_i^{(\ell)} \sigma_i^0
  \biggr) \; ,
\end{equation}
where $\mathcal{N}$ denotes the total number of samples, $\vec{\sigma}^{(\ell)} = (\sigma_1^{(\ell)}, \ldots, \sigma_N^{(\ell)})$ is the $\ell$-th sample, and $\delta(x)\! =\! 1$ for $x\! =\! 0$ and $\delta(x)\! =\! 0$ for $x\! \neq\! 0$. 

\begin{figure}
  \centering
  \subfigure[]{
    \label{fig:RRK10L3overlapA}
    \includegraphics[angle=270,width=0.469\linewidth]{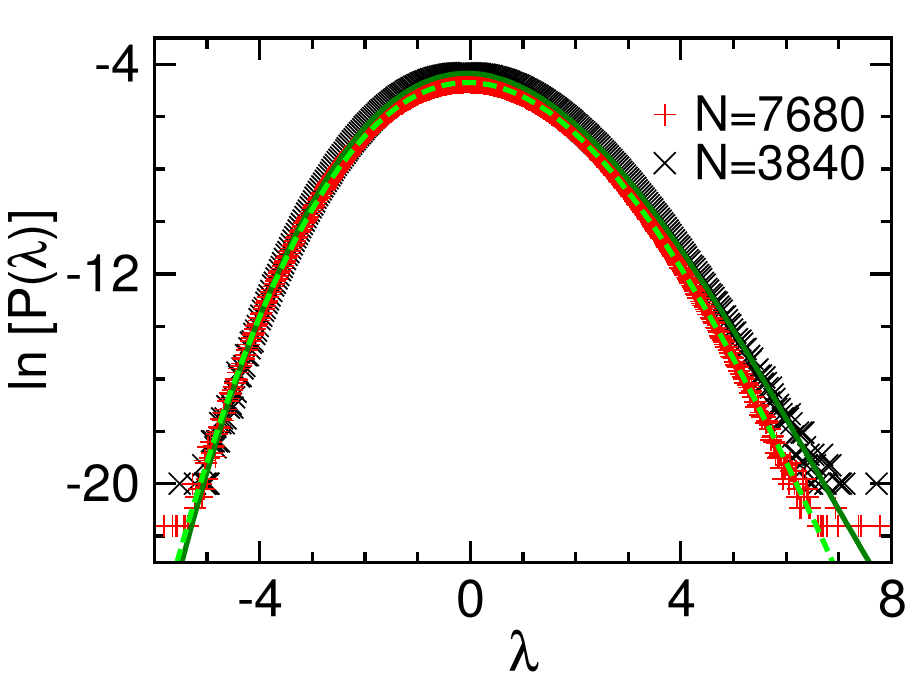}
  }
  \subfigure[]{
    \label{fig:RRK10L3overlapB}
    \includegraphics[angle=270,width=0.469\linewidth]{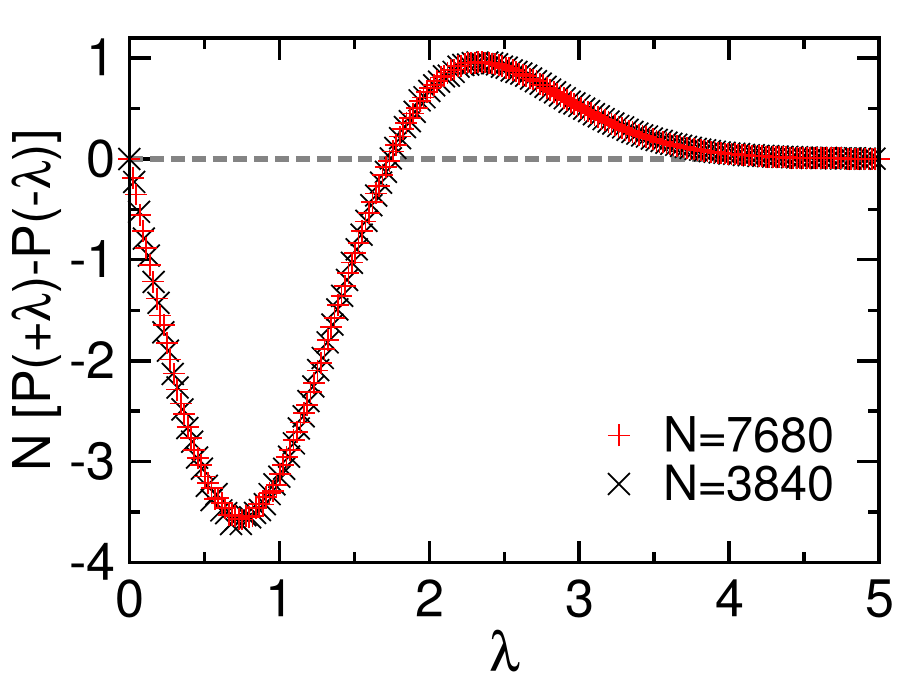}
  }
  \caption{
    \label{fig:RRK10L3overlap}
    The probability distribution function $P(\lambda)$ of the rescaled overlap $\lambda$ is not symmetric for the planted $3$-spin system. The empirical $P(\lambda)$ is obtained by sampling $\mathcal{N} \! = \! 4.8\! \times\! 10^8$ configurations of energy density $u\! = \! -1.61$ through the hybrid-flip MMC algorithm at interval of $100$ sweeps, for a single random graph instance of degree $K \! = \! 10$ (noise $\varepsilon \! = \! 0$). The system size is $N \! = \! 3840$ (crosses) or $7680$ (pluses). (a) The logarithm of $P(\lambda)$ can be fitted by $\ln [P(\lambda)] \! = c_0 \! -\! \lambda^2/2 \! - \! c_1 \lambda \! + \! c_3 \lambda^3$ with $c_1 \! = \! 0.06963$ and $c_3 \! = \! 0.02436$ (solid curve) or with $c_1 \! = \! 0.04939$ and $c_3 \! = \! 0.01766$ (dashed curve). (b) Difference of empirical probability $P(\lambda)$ between positive and negative overlaps of the same magnitude.}
\end{figure}

For the planted $3$-spin systems of degree $K \! \geq \! 4$ at zero noise ($\varepsilon \! = \! 0$), we find that the overlap distribution $P(\lambda)$ is \emph{not} symmetric in $\lambda$, see Fig.~\ref{fig:RRK10L3overlapA} for some representative examples. When the magnitude $|\lambda|$ is small the rescaled overlap $\lambda$ is more likely to be negative than to be positive, while it is more likely to be positive than to be negative as $|\lambda|$ increases beyond certain moderate value [Fig.~\ref{fig:RRK10L3overlapB}]. Overall, the rescaled overlap $\lambda$ is more likely to be negative than to be positive. Our extensive empirical results suggest that the probabilities of $\lambda$ being non-positive ($P_{\leq 0}$) and non-negative ($P_{\geq 0}$) scale with $N$ as
\begin{equation}
  \label{eq:ppn}
  P_{\leq 0} = \frac{1}{2} + \frac{\gamma}{2 \sqrt{N}} \; , \quad
  P_{\geq 0} = \frac{1}{2} - \frac{\gamma}{2 \sqrt{N} } \; ,
\end{equation}
where the asymmetry parameter $\gamma$ is estimated to be $\gamma \approx 1.0$ through the bootstrap method~\cite{Efron-SIAM-1979}. (Very surprisingly, our empirical results on the first moment $\langle \lambda \rangle$ suggest that it is identical to zero.)

We find that the empirical $P(\lambda)$ distribution could be precisely fitted by
\begin{equation}
  \label{eq:Plambda}
  P(\lambda) \propto \exp\Bigl( - \frac{\lambda^2}{2} - U(\lambda) \Bigr) \; ,
\end{equation}
with $U(\lambda)$ being an odd function of the following form
\begin{equation}
  \label{eq:Ulmd}
  U(\lambda) = c_1 \lambda - c_3 \lambda^3 \; .
\end{equation}
Both the linear-order coefficient $c_1$ and the third-order coefficient $c_3$ are positive [Fig.~\ref{fig:RRK10L3overlapA}]. According to the pioneering theoretical work in the 1990s~\cite{VandenBroeck-Reimann-1996,Reimann-VandenBroeck-1996}, the anti-symmetric ``potential energy'' $U(\lambda)$ may make it possible to infer the planted ground state $\vec{\sigma}^0$ by unsupervised learning.

We can formulate an unsupervised learning task for the planted $p$-spin model as follows:  (1) Assume all the observed configurations $\vec{\sigma}^{(\ell)}$ are generated by the following generative model
\begin{equation}
  \label{eq:uspln}
  {\rm Prob}( \vec{\sigma}^{(\ell)} ) \propto
  \exp\Bigl( - U \bigl( \frac{1}{\sqrt{N}} \sum\limits_{i=1}^{N}
  \sigma_i^{(\ell)} \sigma_i^0 \bigr) \Bigr) \; ,
\end{equation}
with $N$ Ising parameters $\{\sigma_i^0\}$. (2) Determine the optimal assignment for this set of model parameters so that these $\mathcal{N}$ observed configurations will be most likely to be observed~\cite{Engel-VanDenBroeck-2001}.

We expect the planted ground state to be the true optimal solution of this unsupervised learning problem. But it is also likely that this problem has many sub-optimal solutions, some of which being the spin glass minimal-energy configurations of Eq.~(\ref{eq:Esigma}). Whether the optimal solution could be efficiently achieved for this machine-learning problem is an open issue and deserves thorough investigations in the future.

Other statistical inference strategies, such as perceptron learning and principal components analysis may also help for the planted $p$-spin model~\cite{Engel-VanDenBroeck-2001,Kabashima-Uda-2004,Braunstein-Zecchina-2006,Huang-Toyoizumi-2015,Zhou-2019b,Hu-etal-2019}. For example, the asymmetric property (\ref{eq:ppn}) of the sampled configurations indicates the existence of an optimal hyperplane separating the $\mathcal{N}$ configuration samples in the most asymmetric manner.

\section{Summary}

In summary, we studied the planted $p$-spin interaction model as an example to explore the feasibility of microcanonical spontaneous symmetry breaking for solving hard inference problems. We confirmed the existence of a discontinuous MSSB phase transition in the regular random graph ensemble [Fig.~\ref{fig:HRRK10L3:b}]. Our numerical results demonstrated that both the canonical route and microcanonical route [Fig.~\ref{fig:TwoRoutes}] could circumvent the low-energy spin glass states to reach the planted ground state, and that the restart strategy is superior to the persistent random walk strategy [Fig.~\ref{fig:WTN1008}]. We also pointed out that inferring the planted ground state for this model system could be projected into an unsupervised learning problem [Fig.~\ref{fig:RRK10L3overlapA}], which may serve as a hard bench-mark problem for various machine-learning algorithms.

The finite-size scaling behavior (\ref{eq:ppn}) and the potential function (\ref{eq:Ulmd}) are quite fascinating and they call for a thorough theoretical understanding. Numerical implementation of the machine-learning strategy (\ref{eq:uspln}) needs to be done in the future. We are also starting to work on other challenging planted ensembles of optimization problems, such as $K$-satisfiability and $Q$-coloring~\cite{Li-Ma-Zhou-2009,Krzakala-Zdeborova-2009,Krzakala-Mezard-Zdeborova-2014}, to get more insights on the MSSB mechanism.

Very recently, two preprints on solving the planted random $3$-spin model of degree $K \! = \! 3$ were posted~\cite{Bellitti-etal-2021,Bernaschi-etal-2021}.  The numerical results of the quasi-greedy algorithm of these two studies are consistent with the physical picture offered in our present work.  To be concrete, we include the theoretical results obtained for this special system in Fig.~\ref{fig:HRRK3L3} of Appendix~\ref{app:mfr}.

\begin{acknowledgments}

One of the authors (H.-J.Z.) thanks Erik Aurell, Yuliang Jin, Federico Ricci-Tersenghi, and Pan Zhang for helpful conversations. This work was supported by the National Natural Science Foundation of China Grants No.~11975295, No.~11947302 and No.~12047503, and the Chinese Academy of Sciences Grant No.~QYZDJ-SSW-SYS018. Numerical simulations were carried out at the Tianwen clusters of ITP-CAS and the Tianhe-2 platform of the National Supercomputer Center in Guangzhou.
\end{acknowledgments}

\begin{appendix}
  \section{Mean field theoretical equations}
  \label{app:mfr}
  
  We employed the replica-symmetric mean field theory of spin glasses to predict the MSSB phase transition for the planted $p$-spin model. To serve as a quick reference, here we list the main theoretical equations of this theory~\cite{Mezard-Parisi-2001,Mezard-Montanari-2006,Matsuda-etal-2011,Zhou-2015}.

    First, the marginal probability $q_i^{\sigma_i}$ of vertex $i$ being in spin state $\sigma_i$ is evaluated through
  \begin{equation}
    q_i^{\sigma_i} =
    \frac{1}{z_i} \prod\limits_{a \in \partial i} p_{a\rightarrow i}^{\sigma_i} \; ,
  \end{equation}
  where $z_i$ is a normalization constant to ensure $q_i^{+1}\! + \! q_i^{-1} \! = \! 1$, and $\partial i$ contains all the clauses that are connected to vertex $i$, and $p_{a\rightarrow i}^{\sigma_i}$ is defined as the probability of vertex $i$ being in spin state $\sigma_i$ if it is affected only by clause $a$ but not by all the other attached clauses.  We define a complementary quantity $q_{i\rightarrow a}^{\sigma_i}$ as the probability of vertex $i$ being in spin state $\sigma_i$ if it is affected by all the attached clauses except clause $a$. In the literature, both $q_{i\rightarrow a}^{\sigma_i}$ and $p_{a\rightarrow i}^{\sigma_i}$ are referred to as cavity probability distributions.

  Let us assume that all the vertices in the set $\partial a$ will be mutually independent if the energetic effect associated with clause $a$ is absent, then the following belief-propagation equation for the cavity probabilities could be derived:
  \begin{subequations}
    \label{eq:bp}
    \begin{align}
      p_{a\rightarrow i}^{\sigma_i} & =
      \frac{1}{z_{a\rightarrow i}} \sum\limits_{\{\sigma_j :
        j \in \partial a\backslash i\}}
      e^{- \beta E_a} \prod\limits_{j\in \partial a\backslash i}
      q_{j\rightarrow a}^{\sigma_j} \; ,
      \label{eq:paibp} \\
      q_{i\rightarrow a}^{\sigma_i} & =
      \frac{1}{z_{i\rightarrow a}} \prod\limits_{b\in \partial i\backslash a}
      p_{b\rightarrow i}^{\sigma_i} \; .
      \label{eq:qiabp}
    \end{align}
  \end{subequations}
  In these expressions, $z_{a\rightarrow i}$ and $z_{i\rightarrow a}$ are two normalization constants, $E_a$ ($\equiv\! -J_a \prod_{j \in \partial a} \sigma_j$) is the energy of clause $a$,  $\partial a$ contains all the vertices which are attached to clause $a$, $\partial a\backslash i$ denotes the residual vertex set obtained by deleting $i$ from $\partial a$, and similarly $\partial i\backslash a$ denotes the residual clause set obtained by deleting clause $a$ from $\partial i$. 

  The total free energy $F(\beta)$ is computed as
  \begin{equation}
    F(\beta) = \sum\limits_{i = 1}^{N} f_i +
    \sum\limits_{a=1}^{M} f_a - \sum\limits_{(i, a)} f_{(i, a)} \; ,
  \end{equation}
  where $f_i$, $f_a$, and $f_{(i, a)}$ are, respectively, the free energy contribution of vertex $i$, clause $a$, and link $(i, a)$ between vertex $i$ and clause $a$. The corresponding expressions are
  \begin{subequations}
    \begin{align}
      f_i & = -\frac{1}{\beta} \ln \Bigl[ \sum\limits_{\sigma_i}
        \prod\limits_{a \in \partial i} p_{a\rightarrow i}^{\sigma_i} 
        \Bigr] \; , \\
      f_a & = - \frac{1}{\beta} \ln \Bigl[
        \sum\limits_{\{\sigma_j : j \in \partial a\}}
        e^{- \beta E_a}
        \prod\limits_{j\in \partial a}
        q_{j\rightarrow a}^{\sigma_j} \Bigr]  \; , \\
      f_{(i, a)} & = - \frac{1}{\beta} \ln \Bigl[
        \sum\limits_{\sigma_i}
        p_{a\rightarrow i}^{\sigma_i} q_{i\rightarrow a}^{\sigma_i} 
        \Bigr] \; .
    \end{align}
  \end{subequations}
  The free energy density is $f \! = \! F / N$. The mean value of the energy density $E(\vec{\sigma})/N$ is evaluated to be
  \begin{equation}
    u = \frac{1}{N} \sum\limits_{a=1}^{M}
    \frac{ \sum\limits_{\{\sigma_j : j \in \partial a\}}
      E_a e^{- \beta E_a}
      \prod\limits_{j\in \partial a}
      q_{j\rightarrow a}^{\sigma_j}
    }{
      \sum\limits_{\{\sigma_j : j \in \partial a\}}
      e^{ - \beta E_a}
      \prod\limits_{j\in \partial a}
      q_{j\rightarrow a}^{\sigma_j}
    } \; .
  \end{equation}
 \begin{figure}
   \centering
   \subfigure[]{
     \label{fig:HRRK4L3:f}
     \includegraphics[angle=270, width=0.469\linewidth]{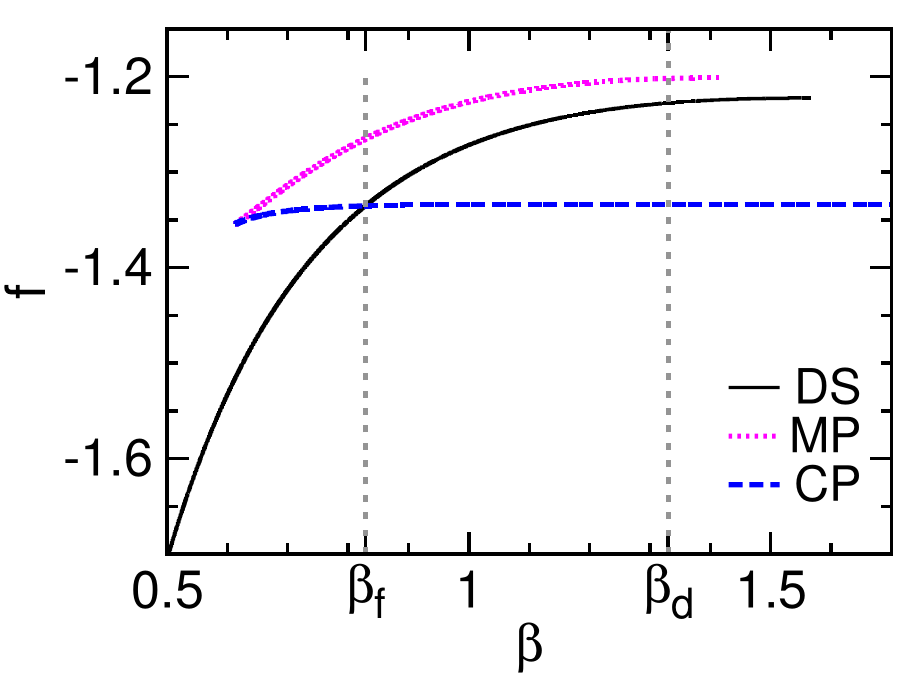}
   }
   \subfigure[]{
     \label{fig:HRRK4L3:s}
     \includegraphics[angle=270, width=0.469\linewidth]{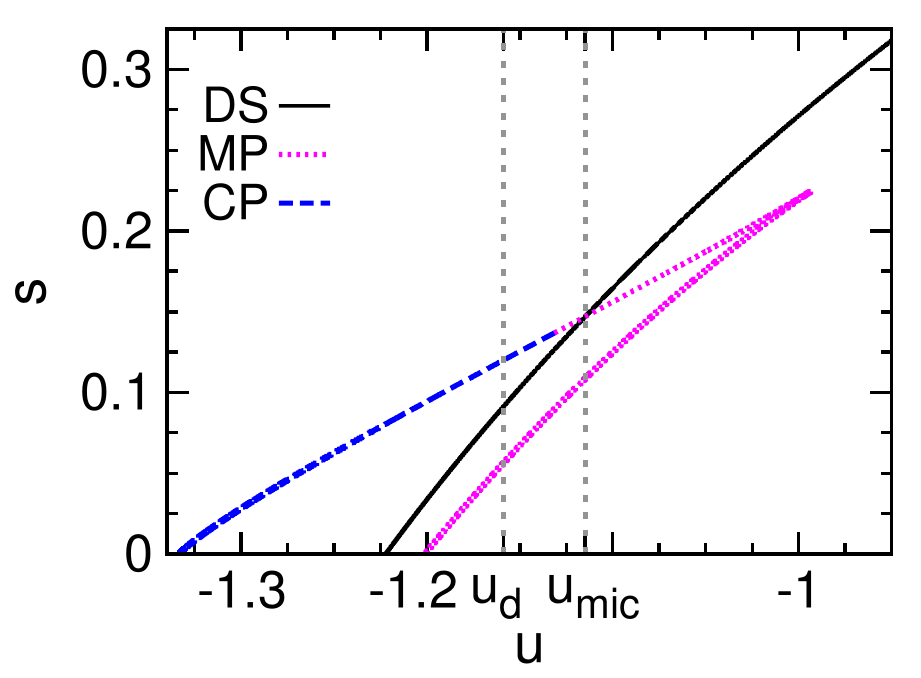}
   }
   
   \subfigure[]{
     \label{fig:HRRK4L3:b}
     \includegraphics[angle=270, width=0.469\linewidth]{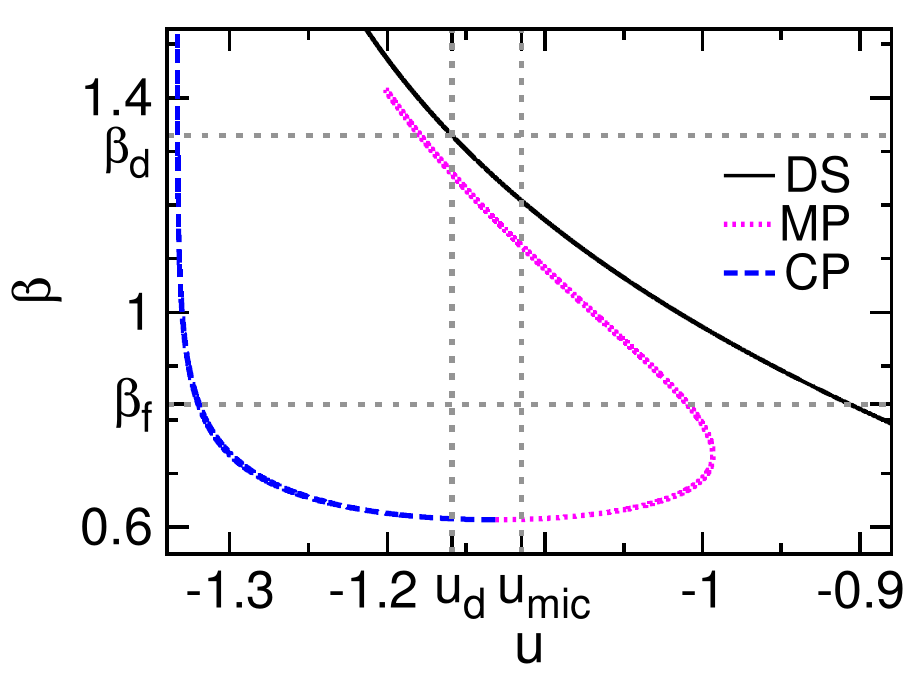}
   }
   \subfigure[]{
     \label{fig:HRRK4L3:m}
     \includegraphics[angle=270, width=0.469\linewidth]{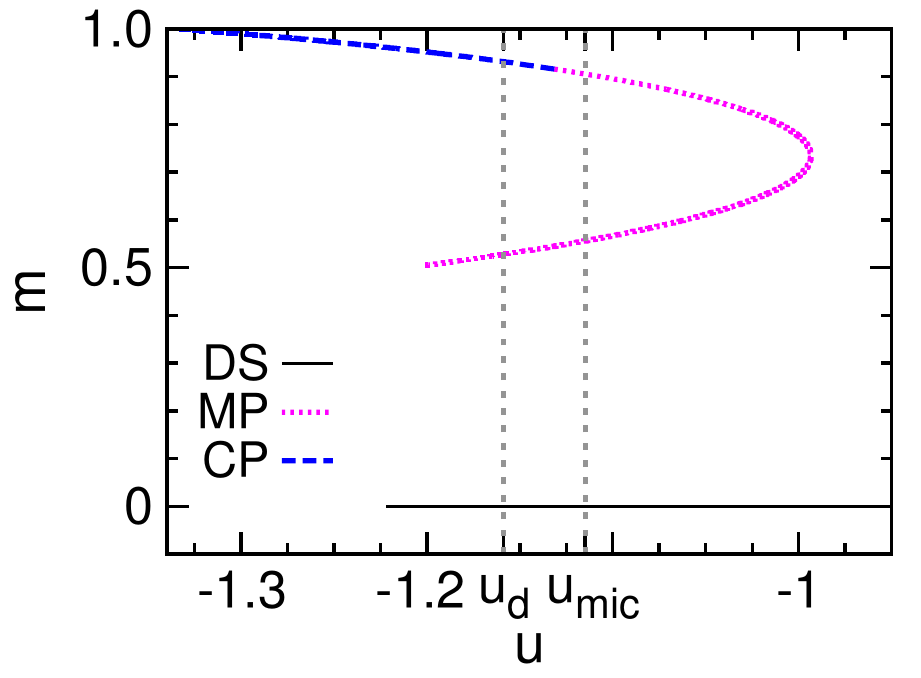}
   }
   \caption{\label{fig:HRRK4L3}
     Same as Fig.~\ref{fig:HRRK10L3}, but for the regular random graph ensemble of vertex degree $K \! = \! 4$ and clause degree $p\! = \! 3$ (noise $\varepsilon \! = \! 0$). (a) Free energy density $f$ versus canonical inverse temperature $\beta$. (b) Entropy density $s$ versus energy density $u$. (c) Microcanonical inverse temperature $\beta$ versus $u$. (d) Mean overlap $m$ versus $u$.
   }
 \end{figure}
 \begin{figure}
   \centering
   \subfigure[]{
     \label{fig:HRRK4L3:f}
     \includegraphics[angle=270, width=0.469\linewidth]{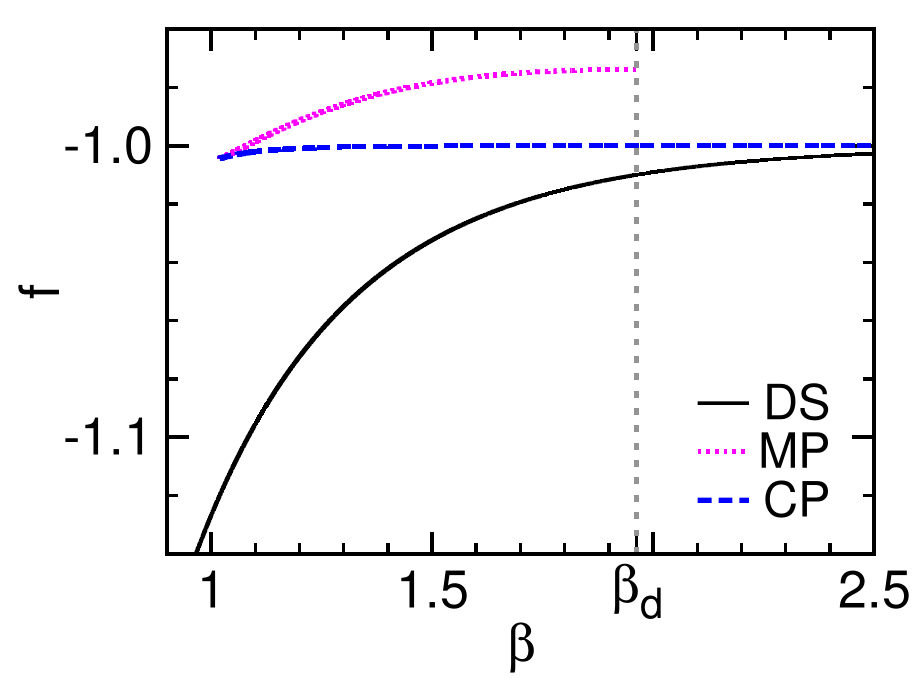}
   }
   \subfigure[]{
     \label{fig:HRRK3L3:s}
     \includegraphics[angle=270, width=0.469\linewidth]{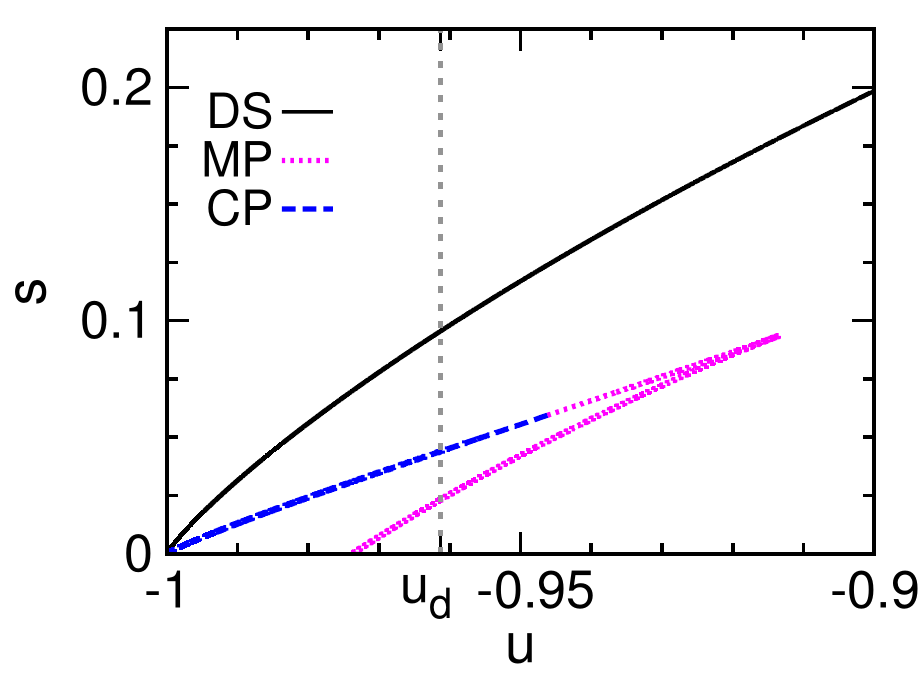}
   }
   
   \subfigure[]{
     \label{fig:HRRK3L3:b}
     \includegraphics[angle=270, width=0.469\linewidth]{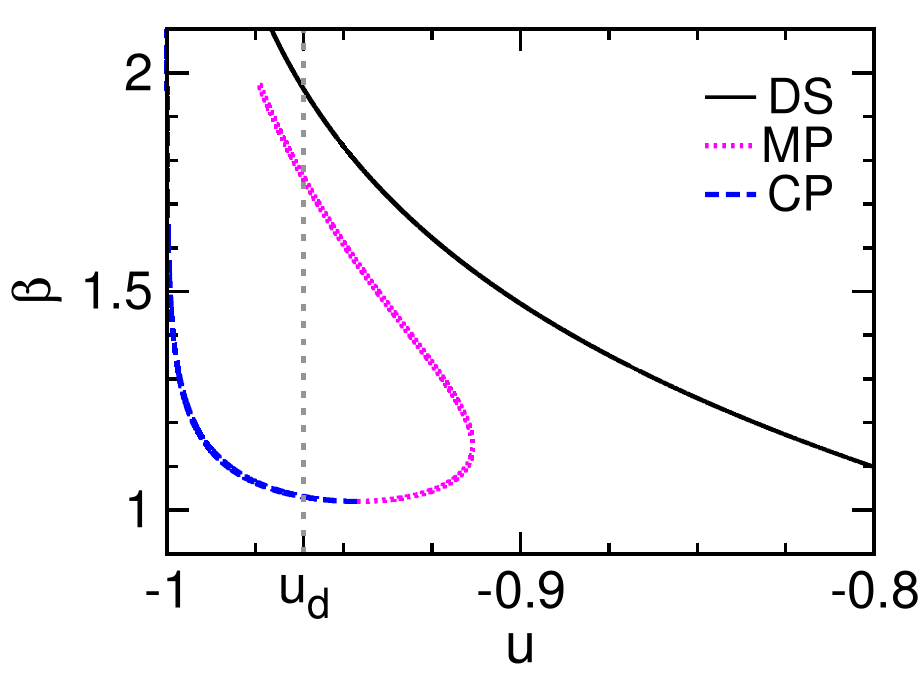}
   }
   \subfigure[]{
     \label{fig:HRRK3L3:m}
     \includegraphics[angle=270, width=0.469\linewidth]{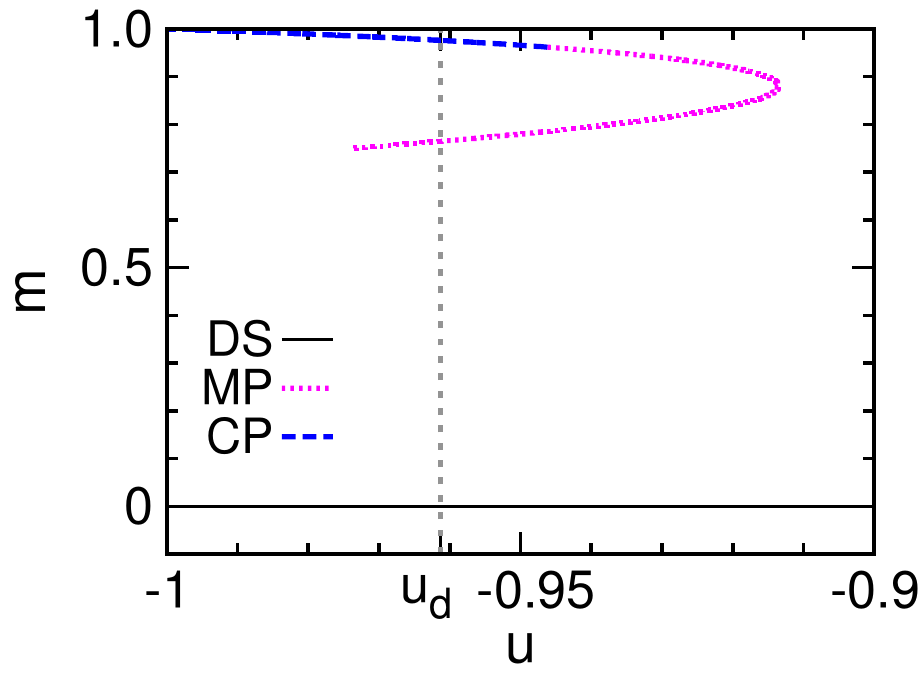}
   }
   \caption{\label{fig:HRRK3L3}
     Same as Figs.~\ref{fig:HRRK10L3} and ~\ref{fig:HRRK4L3}, but for the regular random graph ensemble of vertex degree $K \! = \! 3$ and clause degree $p\! = \! 3$ (noise $\varepsilon \! = \! 0$). (a) Free energy density $f$ versus canonical inverse temperature $\beta$. (b) Entropy density $s$ versus energy density $u$. (c) Microcanonical inverse temperature $\beta$ versus $u$. (d) Mean overlap $m$ versus $u$.
   }
 \end{figure}

 In a regular random graph each clause $a$ is connected to $p$ randomly chosen vertices ($|\partial a| \! = \! p$) and each vertex $i$ is involved in $K$ randomly chosen clauses ($| \partial i| \! = \! K$). For such a uniform system, when the noise level $\varepsilon\! = \! 0$, the cavity probability distribution functions $q_{i\rightarrow a}^{\sigma_i}$ will all be independent of the link indices $(i, a)$ and could be described by a single parameter $\tilde{m}$ as
  \begin{equation}
    q_{i\rightarrow a}^{\sigma_i} = \frac{1+\tilde{m}}{2}
    \delta(\sigma_i - \sigma_i^0)
    + \frac{1-\tilde{m}}{2} \delta(\sigma_i + \sigma_i^0) \; .
  \end{equation}
 The parameter $\tilde{m}$ is the mean value of $\sigma_i \sigma_i^0$ when vertex $i$ is affected only by $K\! - \! 1$ (instead of $K$) attached clauses. The self-consistent equation for $\tilde{m}$ is derived from Eq.~(\ref{eq:bp}) as
 \begin{equation}
   \label{eq:tildem}
    \tilde{m} = \tanh\Bigl[ (K-1)
      \textrm{atanh}\bigl( \tilde{m}^{p-1} \tanh\beta \bigr)
      \Bigr] \; .
  \end{equation}
 The mean value of the overlap $m$ [Eq.~(\ref{eq:mexp})], the mean energy density $u$, and the free energy density $f$,  are computed at a fixed point of Eq.~(\ref{eq:tildem}) as
 \begin{subequations}
   \begin{align}
     m & =   \tanh\Bigl[ K
     \textrm{atanh}\bigl( \tilde{m}^{p-1} \tanh\beta \bigr)
     \Bigr] \; , \\
     u & =   - \frac{K}{p} \Bigl[
       \frac{ \tilde{m}^p + \tanh\beta}
            {1+ \tilde{m}^p \tanh\beta} \Bigr] \; , \\
     f & =  -\frac{1}{\beta} \biggl[
       \ln\Bigl(1+
       \bigl(\frac{1-\tilde{m}^{p-1}\tanh\beta}{1+\tilde{m}^{p-1}\tanh\beta}
       \bigr)^{K}\Bigr) \nonumber \\
       &  \quad \quad \quad  \quad
       + K \ln \Bigl(
       \frac{1+\tilde{m}^{p-1} \tanh\beta}{1+\tilde{m}^{p} \tanh\beta}
       \Bigr) \nonumber \\
       &  \quad \quad \quad \quad
       + \frac{K}{p} \ln \bigl(\cosh \beta + \tilde{m}^{p-1} \sinh \beta \bigr)
       \biggr]
     \; .
   \end{align}
 \end{subequations}
 And the entropy density $s$ is computed following Eq.~(\ref{eq:sval}). The relationship between $s$ and $u$ can be obtained by eliminating $\beta$ from the $u(\beta)$ and $s(\beta)$ curves.

 Equation (\ref{eq:tildem}) always has a stable fixed point $\tilde{m} \! = \! 0$ which corresponds to the disordered symmetric phase with zero overlap ($m\! = \! 0$). As the inverse temperature $\beta$ increases, another stable fixed point appears, which corresponds to the canonical polarized phase with $m$ quite close to unity. These two stable fixed points are separated by an unstable fixed point, which corresponds to the higher-entropy branch of the microcanonical polarized phase (when $\beta$ is relatively small)  or to the lower-entropy branch of the microcanonical polarized phase (when $\beta$ is relatively large).

 The mean-field theoretical results obtained for the planted $3$-body spin model on a regular random network of degree $K \! = \! 4$ are summarized in Fig.~\ref{fig:HRRK4L3}. Qualitatively speaking, these theoretical results are identical to the results shown in Fig.~\ref{fig:HRRK10L3} for $K \! = \! 10$. But the interval between $u_{\rm mic}$ and $u_{\rm d}$ is much smaller and the entropy barrier [defined by Eq.~(\ref{eq:dsexp})] is much higher at $K\! = \! 4$.
 
The planted $3$-body spin model with degree $K\! = \! 3$ is special in that the minimum energy density of the ferromagnetic CP phase is identical to that of the paramagnetic DS phase. We show in Fig.~\ref{fig:HRRK3L3} the mean-field theoretical results obtained for this system. The DS phase experiences a spin glass phase transition at the critical energy density $u_{\rm d} \! = \! -0.961$~\cite{Zhou-2015}. At any fixed energy density $u \! > \! -1$ the entropy density of the DS phase is higher than that of the MP or CP phase, therefore there is no MSSB phase transition in this special system. The MP phase and the CP phase do exist but they are only metastable as compared with the DS phase. The quasi-greedy algorithm of Refs.~\cite{Bellitti-etal-2021,Bernaschi-etal-2021} relaxed the detailed balance condition to ensure that, once the CP phase is reached, the search process will not jump back to the DS phase.

\end{appendix}


%

\end{document}